\documentclass[preprint]{aastex}

\usepackage{graphicx}
\usepackage{natbib}
\bibpunct{(}{)}{;}{a}{}{,}

\newcommand{\bq}{\begin{equation}}
\newcommand{\eq}{\end{equation}}

\newcommand{\simgt}{\lower.5ex\hbox{$\; \buildrel > \over \sim \;$}}
\newcommand{\simlt}{\lower.5ex\hbox{$\; \buildrel < \over \sim \;$}}
\newcommand{\othree}{[O~\small III\normalsize]}

\newcommand{\ntwo}{[N~\small II\normalsize]}

\begin{document}

\title{Tidally Triggered Star Formation in Close Pairs of Galaxies: \\
Major and Minor Interactions}
\author{Deborah Freedman Woods}
\affil{Department of Astronomy, Harvard University 
\\ 60 Garden St., Cambridge, MA 02138}
\email{dwoods@cfa.harvard.edu}

\author{Margaret J. Geller}
\affil{Smithsonian Astrophysical Observatory 
\\ 60 Garden St., Cambridge, MA 02138}

\author{Elizabeth J. Barton}
\affil{Department of Physics and Astronomy, University of California Irvine 
\\ 4154 Frederick Reines Hall, Irvine, CA 92697}

\begin{abstract}
We study star formation in a sample of 345 galaxies in 167 pairs and
compact groups drawn from the original CfA2 Redshift Survey and from a 
follow-up search for companions.  We construct our sample with 
attention to including pairs with luminosity contrast 
$\left | \Delta m_R \right | \geq 2$.   These 57 galaxies with 
$\left | \Delta m_R \right | \geq 2$ provide a set of nearby representative 
cases of minor interactions, a central feature of the hierarchical galaxy 
formation model.  Here we report the redshifts and positions of the 
345 galaxies in our sample, and of 136 galaxies in apparent pairs that 
are superpositions.  In the pairs sample as a whole, there are strong 
correlations between the equivalent width of the H$\alpha$ emission line 
and the projected spatial and line-of-sight velocity separation of the 
pair. For pairs of small luminosity contrast, $\left | \Delta m_R \right | < 2$, 
the member galaxies  show a correlation between the equivalent width of 
H$\alpha$ and the projected spatial separation of the pair.  However, 
for pairs with large luminosity contrast, 
$\left | \Delta m_R \right | \geq 2$, we detect no correlation between 
the equivalent width of H$\alpha$ and the projected spatial separation. 
The relative luminosity of the companion galaxy is more important in 
a gravitational tidal interaction than the intrinsic luminosity of the 
galaxy.  Central star formation across the entire pairs sample depends 
strongly on the luminosity ratio, $\left | \Delta m_R \right |$, a 
reasonable proxy for the mass ratio of the pair; pairs composed of 
similarly luminous galaxies produce the strongest bursts of star formation.  
Pairs with $\left | \Delta m_R \right | \geq 2$ rarely have 
EW(H$\alpha) \gtrsim 70$~\AA.  
\end{abstract}


\section{Introduction} \label{introduction}

Observational studies demonstrate that tidal interactions between galaxies 
trigger enhanced star formation activity.  \citet{larson_tinsley} first 
studied the colors of ``normal'' and ``peculiar'' galaxies to show a 
connection between probable interacting pairs of galaxies and bursts of 
star formation.  Numerous additional studies provide evidence of enhanced 
star formation activity in apparently interacting systems through
measurements of H$\alpha$ emission, galaxy colors, infrared emission, 
and radio continuum emission (e.g.~\citealp{hummel81,kk84,madore86,kenn87,
jonesstein89,seg-wol92,keel93,keel96,liu-kenn95a,liu-kenn95b,donzelli-p97,
bgk,bgk03,lambas03,nikolic04,kauf04}).  Measurements of the equivalent 
width of H$\alpha$ emission (EW(H$\alpha$)) show that high values of 
EW(H$\alpha$) occur preferentially when the projected spatial separation 
of the pair is small \citep{bgk,lambas03,nikolic04}.  \citet{struck05} 
gives a thorough review of galaxy collisions, including the role
of collisions in galaxy evolution, and the influence of the large
scale dynamics on star formation and nuclear activity.

Numerical simulations of major galaxy-galaxy interactions provide a
physical basis for understanding the observations: central bursts of star 
formation result from strong gaseous inflows \citep{mihos_hern96}.  The 
gaseous inflows occur because gravitational tidal torques transfer the 
angular momentum of the gas outward before the final merger.  The 
gravitational tidal torques arise primarily from 
non-axisymmetric structure induced in the galaxy by its companion 
\citep{mihos_hern96}.  The galaxy structure strongly influences the strength 
and timing of the burst of star formation triggered by the gaseous inflows.
Similarly, numerical simulations of minor mergers show that tidal torques
from minor companions provoke non-axisymmetric structure in the main 
disk galaxy \citep{hern_mihos95}.  

Over the history of the universe, galaxy-galaxy interactions link
the process of star formation with the growth of galaxies. According to 
hierarchical structure formation models, these interactions 
play a critical role in the formation and evolution of galaxies
\citep{somerville_prim99,kauffmann99a,kauffmann99b,diaferio99}.  
Simulations show that galaxies grow by accreting other galaxies, most
often minor companions (see the merger tree in \citealt{wechsler02}).  
Encounters between galaxies and minor companions should be the most 
common type of encounter because of the greater fractional abundance 
of low luminosity galaxies. 

The importance of minor interactions in the galaxy formation process 
underscores the importance of examining the process observationally.
However, identifying minor companions observationally is challenging 
because (1) magnitude limited redshift surveys naturally contain relatively 
more pairs of similar magnitude, and (2) directed searches for low-luminosity 
companions around primary galaxies have inherently low success rates 
because of contamination by the more abundant background galaxies.

To examine the relative effects of major and minor encounters on
central star formation, we compile a sample of galaxy pairs 
spanning a wide range of luminosity ratios.  We build on the pairs 
sample of \citet{bgk,bgk03} with targeted observations of systems 
with apparently large luminosity contrasts.  In our final sample
of 167 pairs and compact groups, including 138 with relative photometry, 
$22\%$ of those with photometry have $\left | \Delta m_R \right | \geq 2$.  

Simulations of galaxy-galaxy interactions suggest that the burst of 
central star formation begins at pericentric passage and continues 
for up to several hundred Myr \citep[e.g.][]{mihos_hern94}.  We 
expect the central star formation to decrease as the galaxies 
move farther apart, the burst ages, and the continuum level rises 
\citep{bgk}.  We therefore study the relationship between central 
star formation and the projected spatial separation.  We also
investigate how the luminosity ratio, as a proxy for the mass ratio,
affects this dependence.

In \S\ref{pairs-sample} we specify the selection criteria for our sample.  
\S\ref{observations} contains the observations and data reduction.
We characterize various properties of the sample galaxies in 
\S\ref{character}, including the classification of starbursts and
active galactic nuclei (AGNs), 
the relative magnitude distribution, the absolute magnitude
distribution, and the distribution of H$\alpha$ equivalent widths.
Our results are in \S\ref{results}.  Then in \S\ref{discussion} 
we discuss the results, and  we conclude in \S\ref{conclusion}.

\section{The Pairs Sample} \label{pairs-sample}
\setcounter{footnote}{1}

We assemble a sample of galaxy pairs and compact groups with attention
to including minor companions and satellite galaxies for a study of 
tidally triggered star formation.  The 168 galaxy pairs and compact groups 
that comprise our sample derive from the 786 galaxies in pairs and compact 
groups identified in the CfA2 Redshift Survey (see \citealt{geller-huchra89} 
for a description of the CfA2 Redshift Survey; \citealt{bgk,barton01} 
for the CfA2 pairs sample).  The primary member of each pair or group 
is in Zwicky's Catalogue of Galaxies and Clusters of Galaxies 
\citep{zwicky} and in the Updated Zwicky Catalogue (UZC, \citealt{falco99}).  
The apparent magnitude limit of the UZC, and hence for our primary
galaxies, is $m_{Zw} < 15.5$.  
For one part of our sample, hereafter called the ``EB'' sample, the 
secondary galaxy similarly has a limiting 
magnitude $m_{Zw} < 15.5$; the secondary galaxy is also in the 
UZC.  The EB sample includes 47 pairs and compact groups, which
are part of the photometric sample described in \citet{barton01,bgk03}.  

 We augment the EB sample at large magnitude contrast by a directed 
search for faint companions.  The 120 pairs and compact groups 
identified by our directed search are presented here for the first
time.  The new sample, referred to as the ``DW'' sample, includes
apparent companions to the $m_{Zw} < 15.5$ UZC galaxies that
  were identified by visual inspection of the digitized Palomar Observatory
Sky Survey (POSS) 
E plates by P. Spotts.  Spotts searched the digitized plates
for apparent companions of $m_R \lesssim 16$ within a projected radius of
$\sim50~h^{-1}$~kpc of the Zwicky galaxies with known redshifts.  For a typical 
spiral galaxy with color $B-R = 1.0$, this limit is roughly $m_{Zw} \sim 17$.  
We measure redshifts to eliminate interlopers and thus to identify
pairs with larger magnitude contrast than those typically well-represented 
in magnitude limited redshift surveys.  Magnitude limited 
surveys rarely include pairs with large magnitude contrast because
such pairs reside in the tail of the relative magnitude distribution, 
where there are relatively few galaxies.  It is therefore necessary to 
identify faint companions by a directed search in order to acquire a 
sample of substantial size containing minor interactions.

The pairs that we target here expand the range of luminosity (mass) ratios 
we can use to explore the physics of tidal interactions. For galaxies with 
$m_B \leq -22$ we can observe companions with $10\%$ of the luminosity of 
the primary (similar to the luminosity ratio of the LMC and Milky Way).  
In the DW+EB sample, $21\%$ 
of the pairs have magnitude differences $\left | \Delta m_R \right | \geq 2$.

The pairs must be coincident both in projected spatial separation 
and in recessional velocity, and they must inhabit low density regions.  
We select pairs with a projected spatial separation of 
$\Delta D \leq 55~h^{-1}$~kpc and a line-of-sight velocity separation 
of $\Delta V \leq 1000$~km~s$^{-1}$.  In the case of a compact group, 
in which the number of galaxies is greater than two, each galaxy must meet 
the maximum separation limits when compared to at least one other 
galaxy in the group, not necessarily when compared to all of the 
galaxies in the group, i.e. they satisfy a standard ``friends of friends'' 
algorithm \citep{barton96,huch-gell82}.  Finally, we also require that 
the galaxies have $cz \geq 2300$~km~s$^{-1}$ to limit their 
angular size relative to the size of the  spectrograph slit, 
and to exclude the Virgo cluster.  
 The galaxies in the DW sample all reside 
in low density regions, where the smoothed galaxy number density contrast 
$\rho_{2.5} \leq 2.2$.  The measurement $\rho_{2.5}$ is a density that is smoothed 
over a 2.5~$h^{-1}$~Mpc scale and normalized to the mean survey 
density\footnotemark[1].  Requiring that the galaxies reside in low density 
regions minimizes influence from the surrounding environment \citep{bgk}, 
and thus suppresses effects of the morphology-density relation 
\citep{dressler80}. 
The 47 pairs in the EB sample reside in regions with density contrast 
$\rho_{2.5} \leq 2.7$, slightly higher than the DW sample but still 
low enough to suppress effects of the morphology-density relation.

\footnotetext{
The mean survey density used for the normalization of the
galaxy number density contrast, $\rho_{2.5}$ is calculated
from the CfA2 Redshift Survey galaxy luminosity function, using the
parameters $M^{*} = -18.8, \alpha = -1.0, \phi^* = 0.04$ \citep{marzke94}.
Assuming a cut-off magnitude equal to the faintest absolute magnitude
found in the CfA2 Pairs sample, $M_{cut} = -17.5$, a representative value
of the mean survey density used for the normalization is $n = 0.035$ 
(Mpc/$h)^{-3}$.  The exact value of the mean survey density is 
calculated locally and depends on the galactic extinction correction.
For more details on the galaxy number density contrast, see   
\citet{grogin-geller}. }

The DW sample is intended to augment the EB sample at low luminosity 
and consequently at large magnitude contrast between pair galaxies. 
The EB sample has been analyzed separately in previous papers 
\citep{barton01,bgk03}.
Because the EB sample was originally chosen for emission line rotation 
curve measurements, it has a slight bias toward including galaxies with
H$\alpha$ emission.  Although the EB sample galaxies do not
necessarily contain {\it high} values of H$\alpha$ emission, galaxies
with {\it no} H$\alpha$ emission were preferentially excluded from the 
sample.  Further discussion on the distribution of H$\alpha$ emission 
in the DW and EB samples can be found in \S\ref{dist-ew}.
 The EB sample contains relatively fewer pairs with large luminosity 
contrast compared to the DW sample because the EB sample galaxies are 
magnitude limited; the DW sample galaxies result from our directed search 
for faint companions.  It is necessary to analyze the two samples combined
to demonstrate trends in the data across the wide range
of luminosity contrasts of the systems.   

We obtained medium resolution optical spectra for all 345 of the galaxies 
in the DW+EB samples using the FAST instrument \citep{fast} on the 
1.5~m Tillinghast telescope at the Fred Lawrence Whipple Observatory (FLWO) 
on Mount Hopkins, Arizona.  The 47 pairs and compact groups in the EB sample have 
complete absolute photometry in $B$ and $R$, observed with the 4-shooter
instrument at the FLWO's 1.2~m telescope.  We also 
have complete relative photometry in $B$ and $R$ for 92 of the 121 
pairs and compact groups in the DW sample (\S\ref{phot-sample}),
similarly observed with the 4-shooter instrument at the FLWO's 1.2~m telescope.  
The remaining 29 pairs and compact groups contribute only to the 
spectroscopic analysis.

Spectroscopic  confirmation of apparent companions is important
because many are not coincident in redshift space.  We measured
spectra for 254 galaxies during the years 2003-2004 as part of
our directed search for faint companions for the DW sample.  We 
found 118 coincident in redshift space: $54\%$ of the apparent pairs 
are superpositions.  The superpositions are defined as pairs that do 
not meet our selection criteria of projected spatial separation
$\Delta D \leq 55 h^{-1}$~kpc and line-of-sight velocity separation 
$\Delta V \leq 1000$~km~s$^{-1}$.  The limiting magnitude for
our search is roughly $m_{Zw} \sim 17$ ($m_R \sim 16$ on the
the POSS-E plates).  For the EB sample, because the secondary
galaxies satisfy the UZC magnitude limit $m_{Zw} < 15.5$,
the fraction of superpositions is smaller.
 Table~\ref{pairs-table} lists the position, redshift, 
and EW(H$\alpha$) of the newly identified galaxy pairs and compact 
groups from the CfA2 Redshift Survey, as well as for the previously 
identified EB galaxy pairs and compact groups included in our analysis.  
The galaxy's magnitude difference from its nearest neighbor is indicated 
when available. Table~\ref{isolated} gives the redshift and position of 
the galaxies that are superpositions. 

\begin{figure}[htb!]
\begin{center}
\includegraphics[width=4.5in,angle=90]{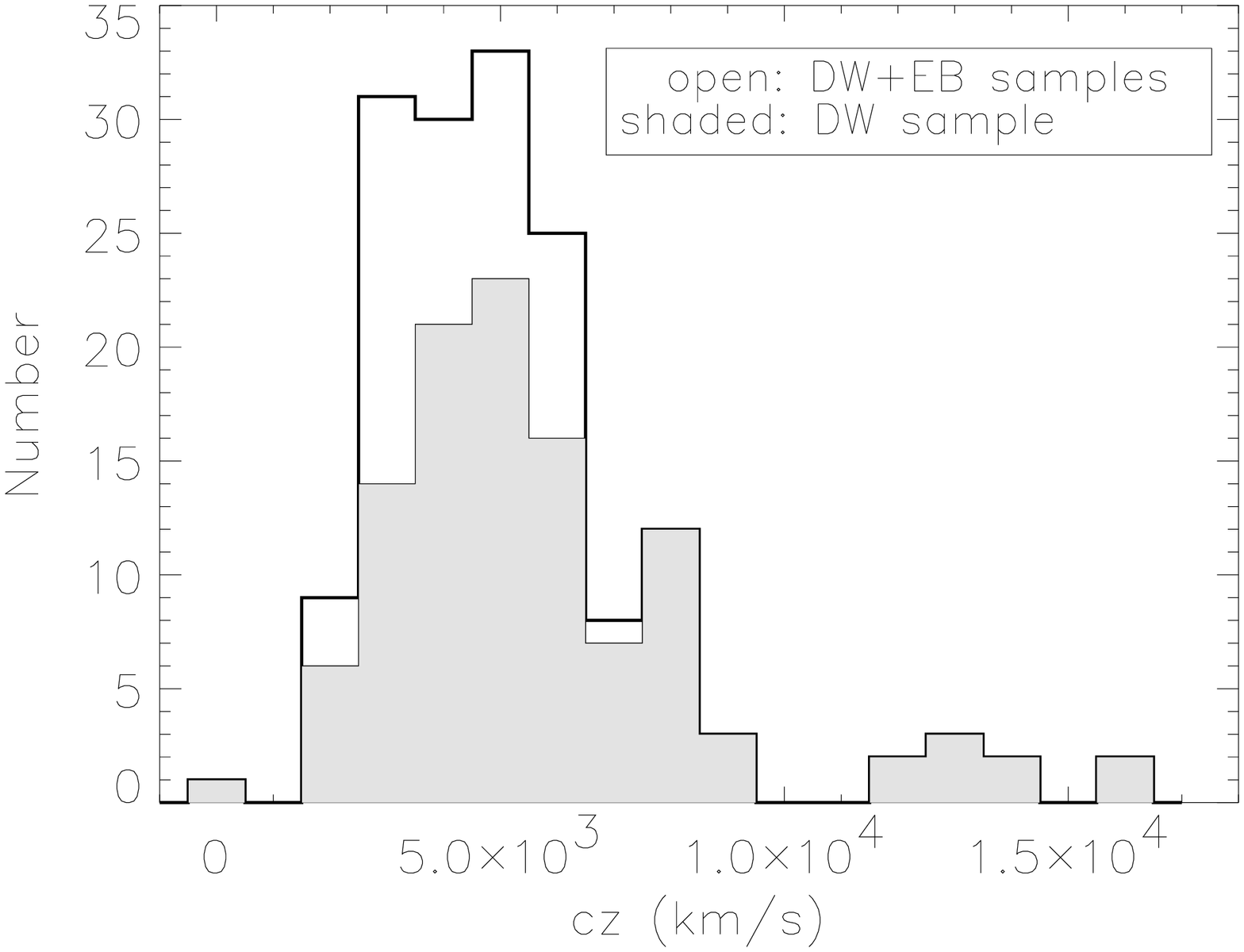}
\caption{The distribution of $cz$, the average recessional velocity 
of the pair.}
\label{hist-cz}
\end{center}
\end{figure}

Fig.~\ref{hist-cz} shows the redshift distribution for the DW+EB galaxies. 
Although the analogous redshift distribution for the UZC galaxies \citep{falco99}
peaks at $cz \sim8500$~km~s$^{-1}$, our selection of low density regions
excludes the Great Wall \citep{geller-huchra89} that covers the range
$cz = 7000 - 10^4$ km s$^{-1}$.

We measure the completeness of the pairs sample by comparing it with the 
original CfA2 North and CfA2 South Surveys.
The range of right ascension and declination for the surveys includes 
$8^h \leq \alpha \leq 17^h$ and $8.^{\circ}5 \leq \delta \leq 44.^{\circ}5$ 
(B1950) for the CfA2 North Survey \citep{geller-huchra89,huchra90,huchra95}, 
and includes $20^h \leq \alpha \leq 4^h$ and 
$-2.^{\circ}5 \leq \delta \leq 48^{\circ}$ for the CfA2 South Survey
\citep{giov-haynes85,giov-haynes89,giov-haynes93,giov86,wegner93,vogeley93}.   
\citet{bgk} estimate that the original CfA2 pair sample of 786 galaxies in 
pairs and compact groups is $70\%$ complete with respect to the UZC 
\citep{falco99}.  The DW+EB sample is complete with respect to the original 
CfA2 South Survey for the 95 objects in 47 pairs and compact groups in low 
density contrast regions $\rho_{2.5} \leq 2.2$ in the right ascension range 
$20^h \leq \alpha \leq 4^h$.  The properties of the sub-sample of 95 
objects are indistinguishable from those of the DW+EB sample as a whole.  
A Kolmogorov-Smirnov (K-S) test of the distributions of the EW(H$\alpha$) 
for the two sub-samples indicates no systematic differences: the 
probability that they are drawn from the same parent sample is $45\%$.
Similarly, the K-S test reveals no systematic differences in the 
distributions of the absolute magnitudes of the galaxies, the relative 
magnitudes of the pairs, and the spatial separations of the galaxies 
from their nearest neighbors in the complete sub-sample compared to the 
entire DW+EB sample.  Hence, the DW+EB sample as a whole is a 
representative subset of the CfA2 South Survey.

\section{Observations and Data Reduction} \label{observations}

To characterize the relative and intrinsic luminosities and the color 
profiles of the galaxies, we obtained photometry in Harris $R$ and $B$ filters.  
We observed 91 of the 120 pairs from the DW sample with the 4-Shooter 
camera mounted on the Fred Lawrence Whipple Observatory's (FLWO's) 
1.2-meter telescope at Mount Hopkins, Arizona, in 2003 March, June, and 
October.  Four out of 11 nights were photometric, enabling 
absolute photometry in B and R for 32 galaxy pairs 
(see Table~\ref{absolute-table}).  We have relative 
photometry in B and R for all 92 pairs.  \S\ref{phot-sample} discusses 
the photometric analysis of the DW sample.  \citet{barton01,bgk03} 
describe the photometric analysis of the EB sample.

To assess the star formation activity of the galaxies, we
measured the EW(H$\alpha$).  We used spectra from the FAST spectrograph 
\citep{fast} mounted on the FLWO's Tillinghast 1.5-meter 
telescope at Mount Hopkins, Arizona, during the years 1994-2004.  Most of 
the spectra for the DW sample are analyzed here for the first time; the 
spectra for the EB sample are included in previous studies 
\citep{bgk,barton01,bgk03}.  \S\ref{spec-sample} describes the analysis 
of the 248 galaxy spectra in the DW sample.

\subsection{Photometric Sample}  \label{phot-sample}

We observed each galaxy pair in Harris R and B filters for a total of
5 minutes in R and 10 minutes in B; in most cases we took two exposures
for each image.  Bias frames, dome flats, sky flats, and dark 
frames comprised our standard calibration data.  When photometric
conditions prevailed, we observed Landolt standard star fields 
\citep{landolt92}.  We used standard imaging data programs from the IRAF  
CCDRED package to reduce the data.  To construct the final image 
for analysis, we first normalized the sky values 
by adding a constant value to every pixel in one of the images to 
equalize the modes of the two images.  Then we combined the images with 
the IRAF task {\it{drizzle}} from the STSDAS  {\it{dither}} package 
(an implementation of the Drizzle algorithm by \citealp{fruchter}.)
Where necessary, we cleaned additional bad pixels from the summed images.

\begin{figure}[htb!]
\begin{center}
\includegraphics[width=4.5in,angle=90]{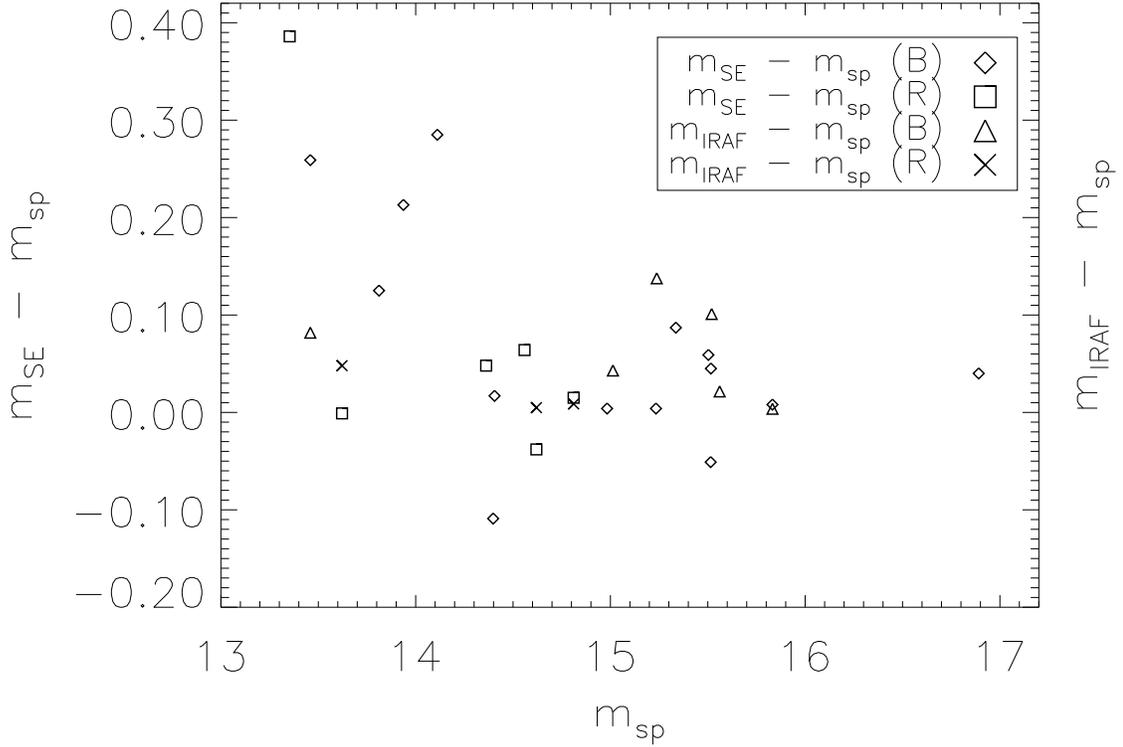}
\caption{Comparisons of $m_{SE} - m_{sp}$ and of $m_{IRAF} - m_{sp}$ 
measured in the same test images.  $m_{SE}$ is the apparent magnitude
measured by SExtractor, a source extraction algorithm \citep{sextractor};
$m_{IRAF}$ is the aperture magnitude measured with IRAF;
and $m_{sp}$ is the apparent magnitude measured by detailed surface
photometry (\citealp{barton01}, Barton, private communication, 2003),
which we take as our reference. }
\label{SEmags}
\end{center}
\end{figure}

We measured galaxy magnitudes with the program SExtractor,
a source extraction algorithm developed by \citet{sextractor}.  
Using detailed surface photometry measurements from the EB sample as 
a standard, we calibrated the SExtractor input parameters to extract 
apparent magnitudes of sources in a set of test images that most closely 
matched the magnitudes determined from detailed surface photometry for 
the same test images \citep{barton01}. For galaxies with apparent 
magnitudes $> 14.2$, there was no 
significant offset between the magnitudes determined by SExtractor 
($m_{SE}$) and by detailed surface photometry ($m_{sp}$); the relative 
scatter is 0.05~mag.  However, for the galaxies with $m_R < 14.2$, the 
SExtractor magnitudes were offset by 0.2~mag, and the relative scatter was 
0.1~mag (see Fig.~\ref{SEmags}).

SExtractor failed to detect one or more galaxies in 16 of 
the pairs when using the calibrated set of input parameters.  The usual 
cause of failure was the apparent location of a galaxy near a bright star 
or in a crowded field.  Modifying the SExtractor input parameters that 
control the background grid size and detection threshold enabled 
detection, but the magnitudes obtained using the modified input parameters 
for SExtractor differed significantly more (by 0.3 mag on average) from 
the ``standard'' detailed surface photometry magnitudes.  Instead, 
IRAF aperture photometry (IRAF task {\it polyphot}) produced much more 
robust results ($m_{IRAF}$) for these objects that compared well with the
detailed surface photometry; the mean offset was 0.07~mag and the scatter 
was 0.1~mag in a sample of 10 test objects.  We extracted aperture 
photometry for every member in a group in both the $B$ and $R$ images 
whenever one galaxy in a group required measurement by aperture 
photometry.  This approach ensures that our measurements of relative 
magnitudes for galaxies within the group were consistent and shared the 
same systematic offsets.

 We determined the
uncertainty in our relative photometry for all of our galaxies,
including the bright ones, using the detailed surface
photometry as a standard.  The standard deviation of the difference
between the SExtractor photometry and the detailed surface photometry
was 0.12~mag, a conservative estimate of our uncertainty.  The uncertainty 
in the IRAF aperture photometry was similar.

\subsection{Spectroscopic Sample} \label{spec-sample}

We acquired long-slit spectra with the FAST spectrograph 
on the FLWO 1.5-meter Tillinghast telescope at Mount Hopkins, Arizona.
The spectra covered wavelengths between 4000 and 7000 \AA\ with a 
dispersion of 300 lines mm$^{-1}$ and a FWHM of 6.2~\AA\ in a 
$3 \arcsec$ wide slit.  The extracted aperture length ranged from 1.7 to 
$32 \arcsec$.  For the galaxy pairs with $2300 < cz < 16,500$~km~s$^{-1}$, 
the aperture covered from 0.42 to 19.5~kpc across the face of the galaxy 
($H_0 = 71$ km s$^{-1}$ Mpc$^{-1}$), with most of the light
coming from the central region of the galaxy. Spectrophotometric standard
stars observed each night provided relative flux calibration.

We measured the EW(H$\alpha$) from the ratio of the line flux to the 
continuum immediately around it.  We used the IRAF task {\it splot} to 
measure the equivalent width of H$\alpha$ and of H$\beta$, [N~\textsc{II}] 
(6583~\AA), and [O~\textsc{III}] (5007~\AA) for diagnostics 
(see \S\ref{isagn}).  We simultaneously fit for 
the [N~\textsc{II}] lines (6548 and 6583 \AA) around H$\alpha$ in case the 
lines were blended.  To account for Balmer absorption around H$\beta$,
we measured the equivalent width of emission in a narrow region  
around the emission line, with the continuum level taken at the base of 
the absorption trough, if an absorption trough was present.  In the case of 
H$\alpha$, the presence of the nearby [N~\textsc{II}] emission lines made it 
difficult to determine the depth of the Balmer absorption trough.  We
adopted values for the  residual absorption corrections measured in the
Nearby Field Galaxy Survey \citep{jansen01,kewley02}, which has similar
spectra and equivalent width measurement techniques. The corrections are $1.5 \pm 0.5$~\AA\ 
and $1.0 \pm 0.5$~\AA\ for EW(H$\alpha$) and EW(H$\beta$), respectively.  
We note that the uncertainty in the Balmer absorption correction is small 
compared to the total uncertainty in our measured equivalent widths, as 
described below. 

We surmised, based on multiple exposures  of the same galaxy, that 
the dominant source of error in equivalent width comes not from the 
method of measuring the emission lines and Balmer absorption, but 
from the repeatability of the slit position for the exposure.  Repeated 
measurements of the EW(H$\alpha$) from multiple exposures of the faintest 
galaxies had an average difference of $18\%$ with an rms scatter of $9\%$, 
giving similar results for galaxies with EW(H$\alpha$) ranging from 5 to 64~\AA\ 
[mean EW(H$\alpha$) = 23~\AA].  We take $18\%$ as a conservative estimate 
of the uncertainty in the measurement of equivalent widths; spectra with 
better signal to noise ratios have a slightly smaller uncertainty.

\section{Characteristics of The Sample} \label{character}

Here we characterize the properties of our pairs sample.  
In \S\ref{isagn} we discuss the spectral classification of the 
objects as starburst galaxies, AGN, or intermediate type galaxies.  We 
then can exclude the objects identified as AGN or intermediate 
classification from the analysis of tidally triggered star formation. 
We thus obtain a  sample of starburst galaxies with spectra dominated 
by photoionization from hot stars \citep[e.g.][]{dopita00,kewley01}, 
not by shocks \citep[e.g.][]{dop_suth95} or by non-thermal or power-law 
continua \citep[e.g.][]{koski78,alexand00}, as is the case for AGN.
The distribution of the EW(H$\alpha$) is described in 
\S\ref{dist-ew}.  In addition to the spectral properties of our 
galaxies (\S\ref{dist-ew}),  we also consider the photometric 
characteristics; \S\ref{magdist} and \S\ref{absmagdist} discuss the 
relative and absolute magnitude distributions, respectively.

\subsection{Starbursts and AGN} \label{isagn}

\begin{figure}[htb!]
\begin{center}
\includegraphics[width=4.5in,angle=90]{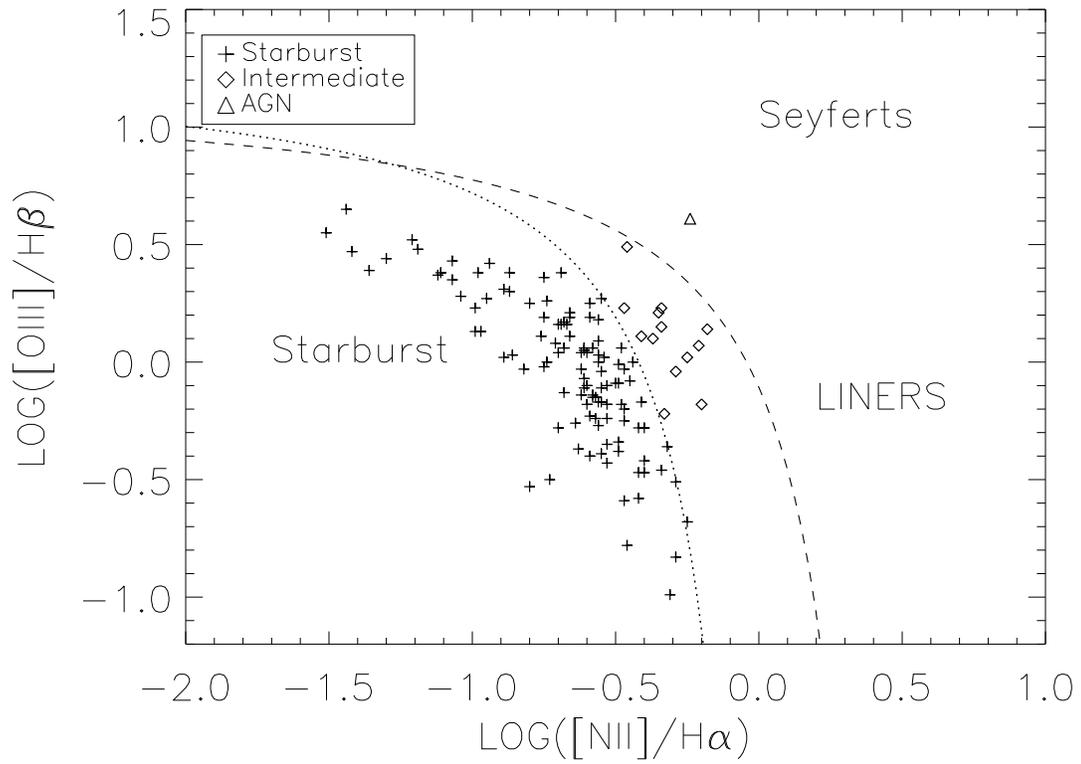}
\caption{Diagnostic diagram showing the identification of starburst galaxies 
and AGN.  The dashed line represents the ionization models of \citet{kewley01}, 
and the dotted line represents the empirical studies of \citet{kauff03c}.}
\label{agnplot}
\end{center}
\end{figure}

Using emission line ratios, we classify galaxies as starburst, AGN, or 
intermediate.  We apply the line diagnostics of \citet{kewley01}, who 
derive an updated classification using the ratios of two sets of emission 
lines \citep[based on the method of][]{bpt,veill-ost87}.  
Using the ratios of \othree~$\lambda$5007/ H$\beta$ and 
\ntwo~$\lambda$6584/ H$\alpha$, we classify the 125 galaxies in 
the DW sample with measurable emission 
in all four lines.  We correct the H$\alpha$ and H$\beta$ lines for
Balmer absorption; we do not correct for reddening because these line 
ratios are nearly independent of reddening.  Fig.~\ref{agnplot} shows the
classifications and the diagnostic line of Kewley et al.  We 
find one AGN above the Kewley et al. diagnostic line, and two more near 
the line.  In addition, we find four objects with 
log(\ntwo~$\lambda$6584/ H$\alpha) > 0$ and \othree~$\lambda$5007 or 
H$\beta$ undetectable.  We count these four objects as AGN.

\citet{kauff03c} argue that the Kewley et al. classification using
\othree~$\lambda$5007/ H$\beta$ and \ntwo~$\lambda$6584/ H$\alpha$
produces a conservative lower limit to the number of AGN in their sample 
of 55,757 Sloan Digital Sky Survey (SDSS) objects with all four lines detected at S/N $> 3$.
Kauffmann et al. define a demarcation line based on their empirical study 
of the positions of the SDSS objects along two apparent branches in the 
 \othree~$\lambda$5007/ H$\beta$ versus \ntwo~$\lambda$6584/ H$\alpha$
diagram.  The Kauffmann et al. classification line includes all objects 
that potentially harbor an AGN, whether or not the AGN dominates the 
optical spectra.  We find 11 more objects above the Kauffmann et al. 
demarcation line, which we exclude from our sample of interacting galaxies.
There are six additional objects in our sample that lie along the Kauffmann 
et al. demarcation line.  We include the six objects on the Kauffmann line in 
our study of interacting galaxies because only a very weak AGN could be 
consistent with the spectra.  The AGN and intermediate galaxies, counting 
even the weakest ones, make up $18\%$ of the objects with all four lines 
measurable, and $9\%$ of the total DW sample. 
In addition, five of the objects in the EB sample were classified
as potential AGN (see \citealt{bgk}).

Our fraction of AGN is a lower limit because we do not classify the 
92 galaxies without measurable emission in one of the lines H$\alpha$ and 
\ntwo~$\lambda$6584, and our cut-off for AGN is conservative when only the 
two lines H$\alpha$ and \ntwo~$\lambda$6584 and not H$\beta$ or 
\othree~$\lambda$5007 are measurable.  For comparison, in
the spectroscopic survey 15R-North, which includes 3149 galaxies and is
$90\%$ complete to R = 15.4,  \citet{carter01} find that $17\%$ of their 
sample has AGN-like emission, and $12\%$ has unclassifiable emission based 
on the classification method of \citet{veill-ost87}.  In a sample of 4921 
galaxies from the SDSS, \citet{miller03} find that at least $\sim 20\%$ of the 
galaxies contain an AGN using the line ratio diagnostics of \citet{veill-ost87} 
and \citet{kewley01}.  Using other methods of classification, 
which include classifying galaxies by the ratio of 
two of the emission lines and applying statistical models, 
Miller et al. find that up to a total of $\simeq 40\%$ of the galaxies 
may contain an AGN. \citet{ho97} find that a higher fraction, $43\%$ of their 
420 emission line galaxies from a sample of 486 galaxies in a nearly 
complete, magnitude limited survey ($B_T \leq 12.5$ mag),  have 
``active'' nuclei (including transition objects), based on a method 
that parallels \citet{veill-ost87}.  The fraction of active galaxies
in our sample is significantly smaller than that in \citet{ho97} 
because the spectrograph slit for Ho et al.'s study is much smaller compared
to the projected size of the galaxy, Ho et al. subtract the stellar
continuum from the spectra, and their spectra have higher signal-to-noise ratio.  
However, our fraction of active galaxies is similar to that of 
\citet{barton01}, who find that 19 out of 150 ($13\%$) of their objects with significant 
H$\alpha$, H$\beta$, [O~\textsc{III}], and [N~\textsc{II}] emission are AGN, 
according to the \citet{veill-ost87} classification method.

\subsection{Distribution of Star Formation Rates} \label{dist-ew}
\setcounter{footnote}{1}

\begin{figure}[htb!]
\begin{center}
\includegraphics[width=4.5in,angle=90]{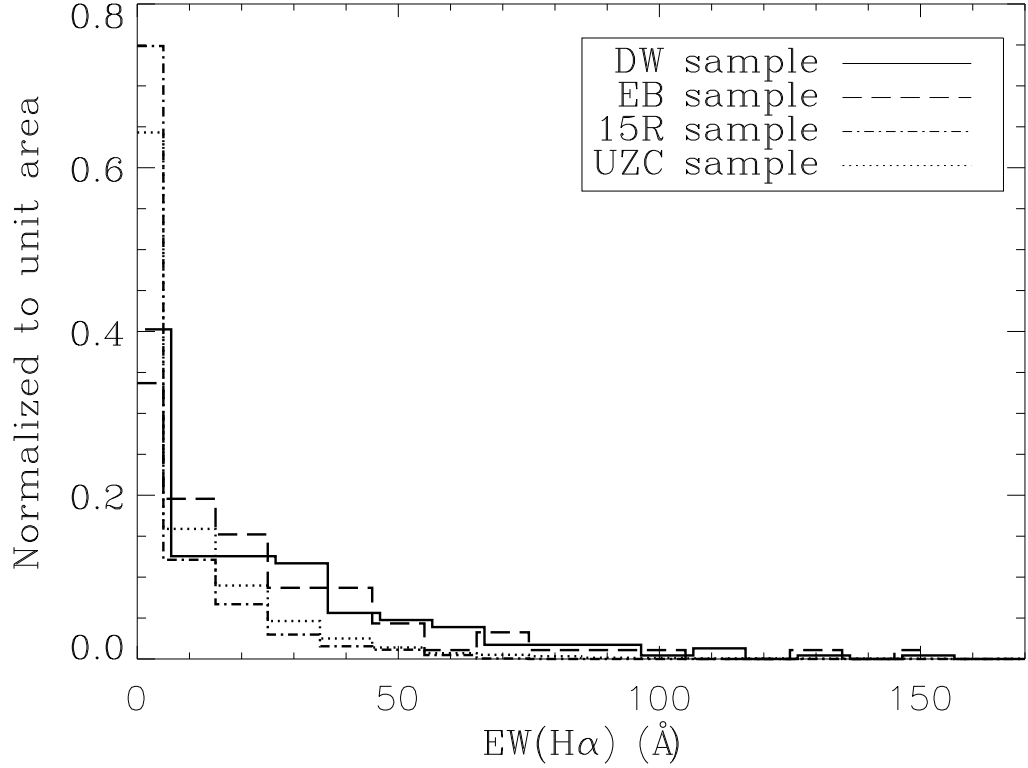}
\caption{Distributions of EW(H$\alpha$) in the DW, EB, 15R-North, 
and UZC samples in the range 0 to 170~\AA.  The distributions are normalized to
unit area.  Galaxies with EW(H$\alpha) > 170$~\AA\ are excluded from 
the figure for the sake of clarity.  There are 2 galaxies in the DW
sample ($0.9\%$), 2 galaxies in the 15R-North sample ($0.1\%$), and 20 
galaxies in the UZC sample ($0.2\%$) with EW(H$\alpha) > 170$~\AA.}
\label{ew4samples}
\end{center}
\end{figure}

The equivalent width of the H$\alpha$ emission line measures a 
combination of starburst age and strength.  We compare the 
distribution of the EW(H$\alpha$) in the DW sample  %\footnotemark[1]
 to a ``field'' galaxy sample, 15R-North 
\citep{carter01}. The 15R-North Survey is a complete, uniform, 
magnitude limited ($R \leq 15.4$) spectroscopic survey.  The
galaxies for the 15R-North Survey were selected from the POSS I E
plates, and spectra were measured on the FAST instrument using
a slit width of 3\arcsec and a slit length of 3\arcmin.  Thus,
the methodology of the 15R-North Survey and the DW sample are in
excellent agreement: the galaxies are selected from the same
 set of plates and measured on the same instrument using the same 
slit width.  Fig.~\ref{ew4samples} shows the distributions of
EW(H$\alpha$) in the DW, EB, 15R-North, and UZC samples.

There is an excess of moderate to high H$\alpha$ emission in the
DW sample compared to a selection of all 15R-North galaxies that fall 
into the same redshift range and apparent magnitude range as our sample 
($2300 < cz < 16,500$~km~s$^{-1}$, and $m_R \lesssim 17$).
As shown in Table~\ref{comp-table},
we measure EW(H$\alpha) > 10$~\AA, representing at least mild star 
formation activity, for $59\%$ (136 out of 230) of the galaxies in the DW 
sample.  In the 15R-North galaxies within the selected redshift
range and apparent magnitude range, $25\%$ (421 out of 1675) have
EW(H$\alpha) > 10$~\AA.
The distribution of EW(H$\alpha$) of our DW sample also shows a 
significantly larger fraction of galaxies with high equivalent widths 
than are present in the 15R-North sample.  Of the 15R-North galaxies
in the same redshift range and apparent magnitude range as our sample,
vigorous star formation, e.g. EW(H$\alpha) > 70$~\AA, is
present in $0.2\%$ (3 out of 1675) of the galaxies.  In the DW sample,  
$9\%$ (20 out of 230) of the galaxies have EW(H$\alpha$) $> 70$~\AA.
The comparison suggests that a selection favoring close pairs 
biases the EW(H$\alpha$) distribution toward higher values, as expected 
if there is a physical connection between the interaction and star
formation.  We note that the 15R-North Survey includes 298 galaxies
in pairs and n-tuples that satisfy our selection criteria: projected
spatial separation $\Delta D < 50~h^{-1}$~kpc, line-of-sight velocity 
separation $\Delta V < 1000$~km~s$^{-1}$, and inhabit the same range
$2300 < cz < 16,500$~km~s$^{-1}$.  Some of the highest values of
EW(H$\alpha$) in the 15R-North sample actually occur for galaxies in 
pairs.  If the pair galaxies in the 15-R North sample were excluded,
then the difference in the distributions of EW(H$\alpha$) in
the 15-R North sample and the DW sample would be even more 
pronounced.  Despite the inclusion of some pairs in the 15-R North 
sample, we see that the DW sample, which is composed entirely of pairs 
and $n$-tuples, preferentially includes high values of EW(H$\alpha$) 
compared to the ``field'' galaxy sample 15R-North.

\begin{figure}[htb!]
\begin{center}
\includegraphics[width=4.5in,angle=90]{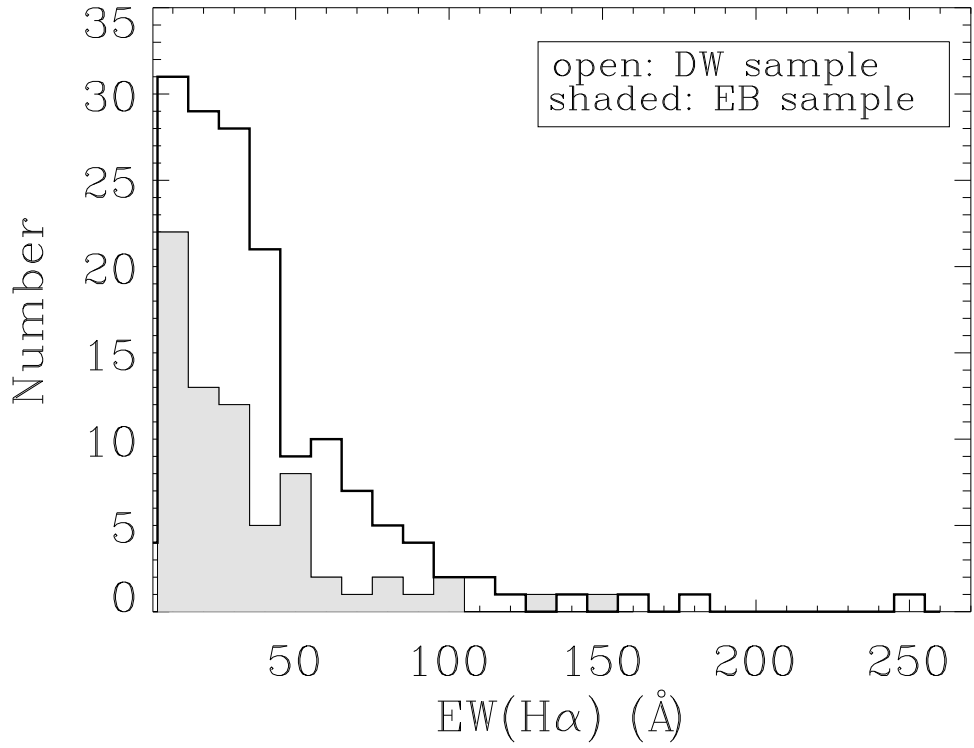}
\caption{Distributions of EW(H$\alpha$) in the DW and EB samples
for the galaxies with EW(H$\alpha) \geq 3.5$~\AA\ (after correction for
Balmer absorption).  Not shown are the EW(H$\alpha) < 3.5$~\AA\ galaxies, which 
make up $32\%$ (74/230) of the DW sample and $24\%$ (22/92) of the EB sample.}
\label{hist-ewplot}
\end{center}
\end{figure}

We also note that the DW and EB samples have remarkably similar distributions 
of EW(H$\alpha$) (see Fig.~\ref{hist-ewplot}).  There is a very slight increase 
in the fraction of EB galaxies with EW(H$\alpha) > 10$~\AA\ compared to that 
in the DW sample, possibly the result of the EB galaxies' bias toward non-zero 
H$\alpha$ emission or more likely due to both the primary and the secondary
galaxies' selection in $B$, whereas the DW primaries are selected in $B$
and the secondaries in $R$.
However, the fraction of galaxies with {\it moderate} to {\it high} 
values of H$\alpha$ emission are identical in the DW and EB samples:
$23\%$ of the DW and EB galaxies have EW(H$\alpha) > 40$~\AA,
and $9\%$ have EW(H$\alpha) > 70$~\AA.  The moderate to high
values of EW(H$\alpha$)  define the envelope of any measurable
correlation between EW(H$\alpha$) and $\Delta D$ or 
$\left | \Delta m_R \right |$ because they represent galaxies
with available gas and orbital positions conducive to triggered
star formation.

\subsection{Relative Magnitude Distribution} \label{magdist}

\begin{figure}[htb!]
\begin{center}
\includegraphics[width=4.5in,angle=90]{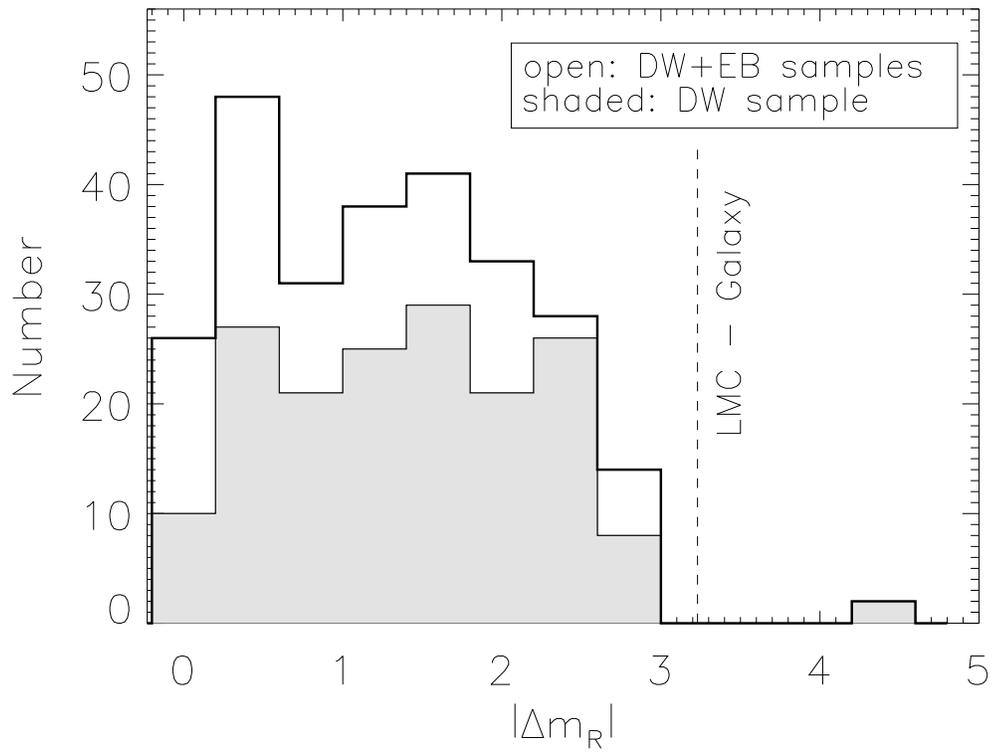}
\caption{Distribution of $\left | \Delta m_R \right |$ for the galaxy 
sample.  The uncertainty in $\left | \Delta m_R \right |$ is 
$\pm 0.17$~mag.  The dashed line indicates $\left | \Delta m_R \right |$ 
between our Galaxy and the LMC \citep{weinberg00}.}  
\label{magdifr-both}
\end{center}
\end{figure}

The magnitude differences  for the galaxy pairs in the combined DW+EB 
sample cover the range  $0 \leq \left | \Delta m_R \right | \leq 4.4$; 
there are 57 galaxies with $\left | \Delta m_R \right | \geq 2$. 
We measured magnitude differences in the $R$ band for $81\%$ (260 out of 322) 
of the galaxies in our spectroscopic sample with available photometry.  We 
use the $R$ magnitude difference between pair galaxies as a proxy for 
the galaxy mass ratio.  We use the magnitude differences only to separate 
our sample into two coarse bins, $\left | \Delta m_R \right | < 2$ and
$\left | \Delta m_R \right | \geq 2$.
Fig.~\ref{magdifr-both} shows the distribution of magnitude 
differences for all of the galaxies, excluding AGN or objects of 
intermediate classification.  

Pairs with $\left | \Delta m_R \right | \geq 2$ are particularly interesting 
because (1) minor interactions are more common than major interactions because 
minor companions are far more common, (2) hierarchical models of galaxy 
formation show that minor mergers occur frequently in a galaxy's history, and
(3) the impact of minor interactions on star formation is not well understood
from either an observational or theoretical perspective.  Therefore, studying the 
effects of minor interactions is crucial to understanding galaxy formation.

\subsection{Absolute Magnitude Distribution} \label{absmagdist}

We use the UZC magnitudes to estimate the absolute magnitude 
for all of the galaxies in our sample with photometry; the UZC contains the 
apparent magnitude, $m_{Zw}$,  of the primary galaxy and its redshift. 
We determine the apparent and absolute magnitude of the secondary galaxy 
from its B magnitude relative to the primary.  Because all of our absolute 
magnitudes are based on the UZC  magnitudes, all of our galaxies 
share any systematic offsets present in the UZC.

\begin{figure}[htb!]
\begin{center}
\includegraphics[width=4.5in,angle=90]{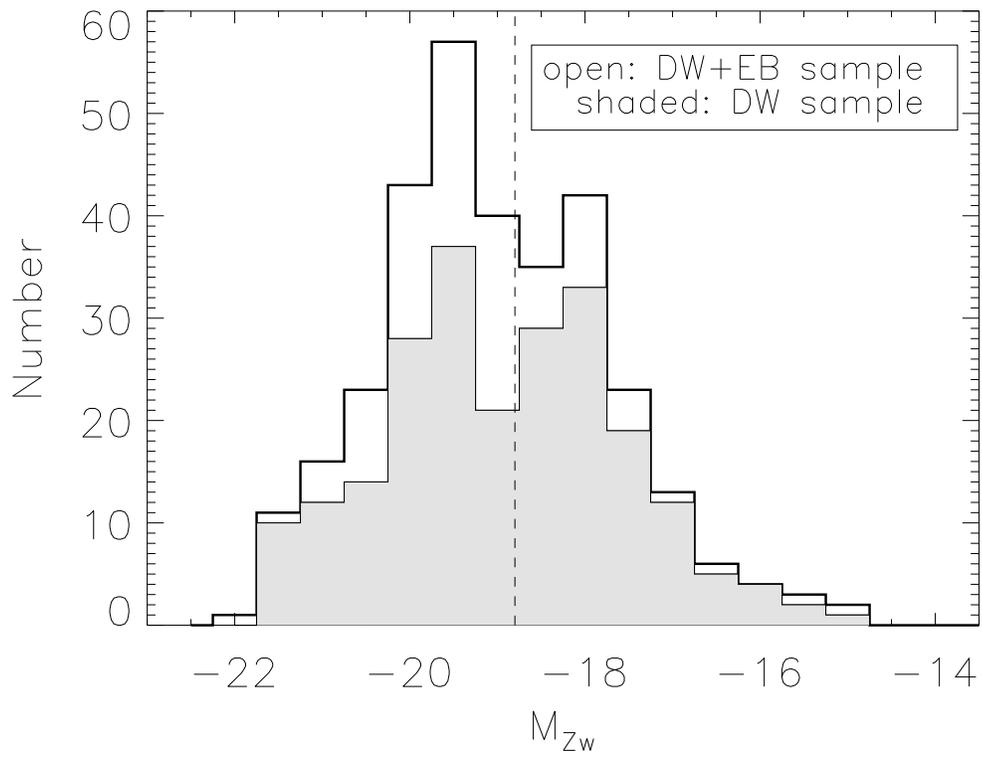}
\caption{Distribution of $M_{Zw}$ in the DW+EB sample.  The dashed
line shows the value of $M_{Zw}^{*} = -18.8$ \citep{marzke94}
for the CfA2 Redshift Survey.}
\label{DWEBabszwic}  
\end{center}
\end{figure}

We compare our absolute $B$ photometry with the results from the UZC.  
The difference in magnitude, $m_{B,DW} - m_{Zw}$, for the 34 galaxies 
with $m_{Zw} \geq 15.0$ observed under photometric conditions shows a
mean offset of $<m_{B,DW} - m_{Zw}> = 0.32$ mag, where $m_{Zw}$ generally 
overestimates the apparent magnitude of the galaxy.  The most likely reason 
for the offset is the fainter limiting isophote of our 
measurements.  The standard deviation, $\sigma_{m_{B,DW} - m_{Zw}} = 0.43$~mag.  
  Our comparison with the Zwicky photometry is consistent
with \citet{bothun}, who find that $m_{Zw}$ corresponds well to 
$b_{26}$ with a scatter of 0.31~mag, in their study of 107 cluster spirals,
including 66 galaxies with $m_{Zw} > 14.9$.  \citet{grogin-geller99}
also find a similar scatter, 0.32~mag, between $m_{Zw}$ and their photometric
analysis of 230 galaxies with $m_{Zw} \leq 15.5$ and eight galaxies with
$15.6 \leq m_{Zw} \leq 15.7$.  In addition, Grogin \& Geller find a negligible 
bias in the absolute Zwicky magnitudes computed from the apparent Zwicky
magnitudes and the UZC redshifts, compared to their absolute magnitudes in the 
range $-20.5 \lesssim M_{Zw} \lesssim -18$.  We conclude that the Zwicky
catalog provides satisfactory estimates of the calibrated apparent magnitude, 
and we use them for all of our galaxies where applicable, in order
to maintain internal consistency.  Fig.~\ref{DWEBabszwic} shows the 
distribution of absolute magnitudes for the DW+EB samples.

\section{Results} \label{results}

Our goal is to isolate the observable properties of the galaxy
or its companion that influence the star formation over the course of the
interaction.   Numerical simulations of tidally triggered star formation 
\citep[e.g.][]{mihos_hern94,mayer01} predict a correlation between
the star formation rate and the time since pericentric passage,
which is related to the spatial separation of the galaxies.
\citet{bgk} discovered a correlation between EW(H$\alpha$) and 
the projected spatial separation $\Delta D$, and between EW(H$\alpha$) 
and the line-of-sight velocity separation $\Delta V$ observationally;
studies by \citet{lambas03} and \citet{nikolic04} confirmed the correlation
between galaxy emission line properties and $\Delta D$ or $\Delta V$.
Here, we extend the observations to pairs with a larger range of relative 
luminosities than previously explored.  

We investigate whether the luminosity contrast of the pair is important 
for determining the effectiveness of the tidal interaction.
In their study of pair galaxies in the Two Degree Field (2dF) Survey, \citet{lambas03} 
find that the star formation activity in pair galaxies in the 2dF 
Survey depends on the relative luminosity of the pair.  In contrast, 
for SDSS close pairs with $\left | \Delta m_z \right | < 2$,
\citet{nikolic04} determine that the star formation rate shows no
 dependence on the luminosity or morphological type of the
companion galaxy.  However, Nikolic et al. suggest that their
results can be reconciled with those of Lambas et al. because 
the distributions of the luminosity contrast of their samples differ.

We use our sample with $\left | \Delta m_R \right | \geq 2$ 
to begin to disentangle the influence of intrinsic galaxy properties 
from the influence of relative properties of the pair.  In order
not to confuse intrinsic galaxy properties with effects of the 
interaction, we separate the sample by intrinsic luminosity and 
test for evidence of tidally triggered star formation in each
sub-sample.  Intrinsically low luminosity galaxies 
tend to have younger stellar populations and contain more gas
and less dust than intrinsically luminous galaxies, and hence
larger values of EW(H$\alpha$) independent of tidal interaction 
with another galaxy.  These generally lower mass galaxies are 
also more strongly affected by supernova triggered star formation 
\citep{lada78,elmegreen95,boss03}, which would not correlate with
$\Delta D$. 

Galaxy morphology may also be a factor in determining the effectiveness 
of the tidally triggered star formation, but morphological classification 
is beyond the scope of this paper.  We thus note that 
elliptical galaxies and some early spiral galaxies, which have less gas and 
dust, would show smaller EW(H$\alpha$) at every 
$\Delta D$.  The inclusion of elliptical galaxies in our sample  
weakens correlations between EW(H$\alpha$) and $\Delta D$ 
relative to a sample containing exclusively gas-rich late 
spirals and irregular galaxies.  The gas-rich galaxies in our sample 
should, however, define the envelope of a measurable correlation between 
EW(H$\alpha$) and $\Delta D$, if such a correlation prevails.

In \S\ref{whole}, we examine
the correlation between  EW(H$\alpha$) and $\Delta D$, and between 
EW(H$\alpha$) and $\Delta V$ for the sample as a whole.  We consider 
the star formation activity in sub-samples of intrinsic luminosity in 
\S\ref{absolutemag}.  In \S\ref{res-rel}, we test the correlation 
between EW(H$\alpha$) and $\Delta D$ for sub-samples selected by relative 
luminosity, paying particular attention to the pairs with
$\left | \Delta m_R \right | \geq 2$.  We also consider the relationship 
between  EW(H$\alpha$) and $\Delta m_R$.

\subsection{EW(H$\alpha$) in the Sample as a Whole} 
\label{whole}

\begin{figure}[htb!]
\begin{center}
\includegraphics[width=4.5in,angle=90]{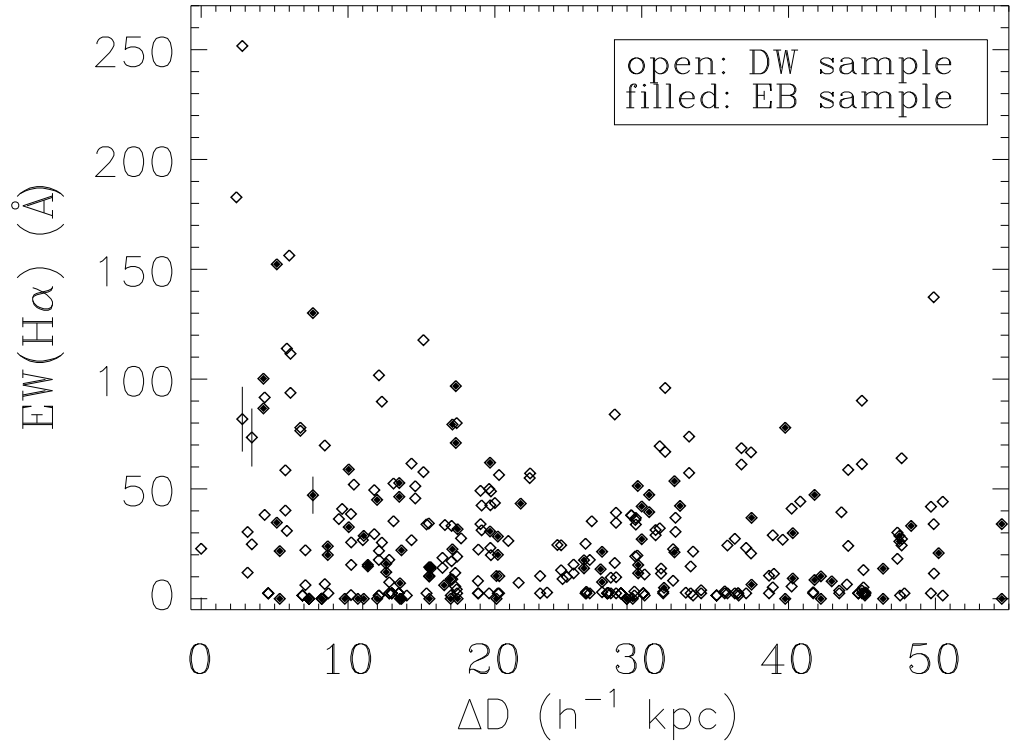}
\caption{$\Delta D$ versus EW(H$\alpha$) for the 322 galaxies  
in the DW+EB sample.  $\Delta D$ is the projected spatial
separation to the nearest neighbor.  Representative error bars
show the measurement uncertainty of $\pm 18\%$ for EW(H$\alpha$).}
\label{ew-sepg}   
\end{center}
\end{figure}

Fig.~\ref{ew-sepg} shows EW(H$\alpha$) versus $\Delta D$ for the 322 galaxies
in the DW+EB sample, excluding known AGN.  Here we include all galaxies with 
FAST spectra, whether or not 1.2~m photometry is available.
The EW(H$\alpha$) is correlated with $\Delta D$ in
the sample as a whole.  A Spearman rank correlation test of EW(H$\alpha$) 
and $\Delta D$ produces a correlation coefficient, $C_{SR}$, of -0.14,
indicating that EW(H$\alpha$) and $\Delta D$ are anti-correlated.
The probability of no correlation, $P_{SR}$, is $8.2\times 10^{-3}$. 
Almost all of the largest EW(H$\alpha$) occur for galaxies with small 
projected spatial separation.  At projected spatial separations 
$\gtrsim 20~h^{-1}$~kpc, very few galaxies in the sample show large 
EW(H$\alpha$).

\begin{figure}[htb!]
\begin{center}
\includegraphics[width=4.5in,angle=90]{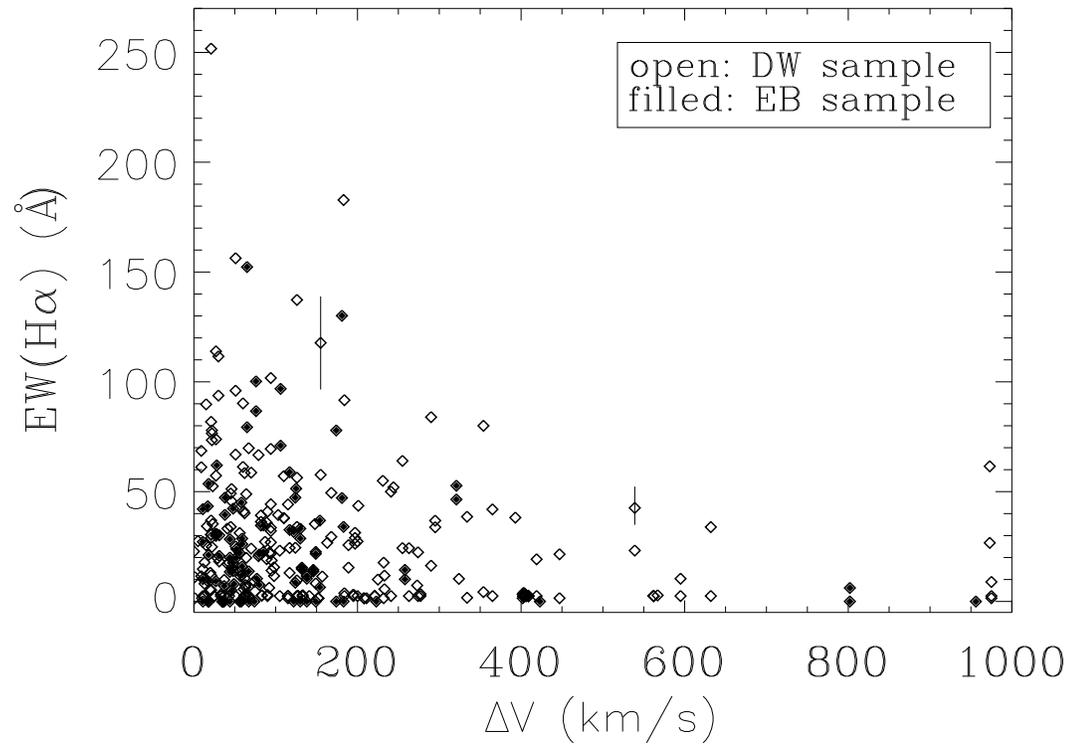}
\caption{ $\Delta V$  versus EW(H$\alpha$) for the 322 galaxies  
in the DW+EB sample. $\Delta V$ is the line-of-sight
velocity separation to the nearest neighbor. Representative error
bars are $\pm 18\%$ in EW(H$\alpha$).}  
\label{ew-veldif}
\end{center}
\end{figure}

The EW(H$\alpha$) is also correlated with $\Delta V$  
(Fig.~\ref{ew-veldif}).  A Spearman rank correlation test between 
EW(H$\alpha$) and $\Delta V$ for the 322 galaxies in the DW+EB samples
 produces $C_{SR} = -0.16$, with $P_{SR} = 3.3 \times 10^{-3}$.  Galaxies
with large EW(H$\alpha$), indicating strong or recent bursts of star 
formation, have smaller relative velocities.  We thus confirm the
results of \citet{bgk}.

\subsection{Intrinsic Luminosity and EW(H$\alpha$)}
\label{absolutemag}

We examine the effect of the intrinsic luminosity of the galaxies on 
the correlation between EW(H$\alpha$) and $\Delta D$. 
Here we include only the 260 galaxies with FLWO 1.2~m photometry, 
all of which have FAST spectra.  We divide the galaxies into subsets
by absolute magnitude, where the intrinsically luminous galaxies with 
$M_{Zw} < M_{Zw}^{*}$ ($M_{Zw}^{*} = -18.8$, assuming $h=1$ 
\citealp{marzke94})  are considered separately from the low
luminosity galaxies with $M_{Zw} > M_{Zw}^{*}$.  We limit 
$\left | \Delta m_R \right | < 2$ for both the intrinsically bright 
and faint subsets to minimize the influence of the luminosity contrast 
when studying the galaxy properties as a function of intrinsic luminosity.

\begin{figure}[htb!]
\begin{center}
\includegraphics[width=4.5in,angle=90]{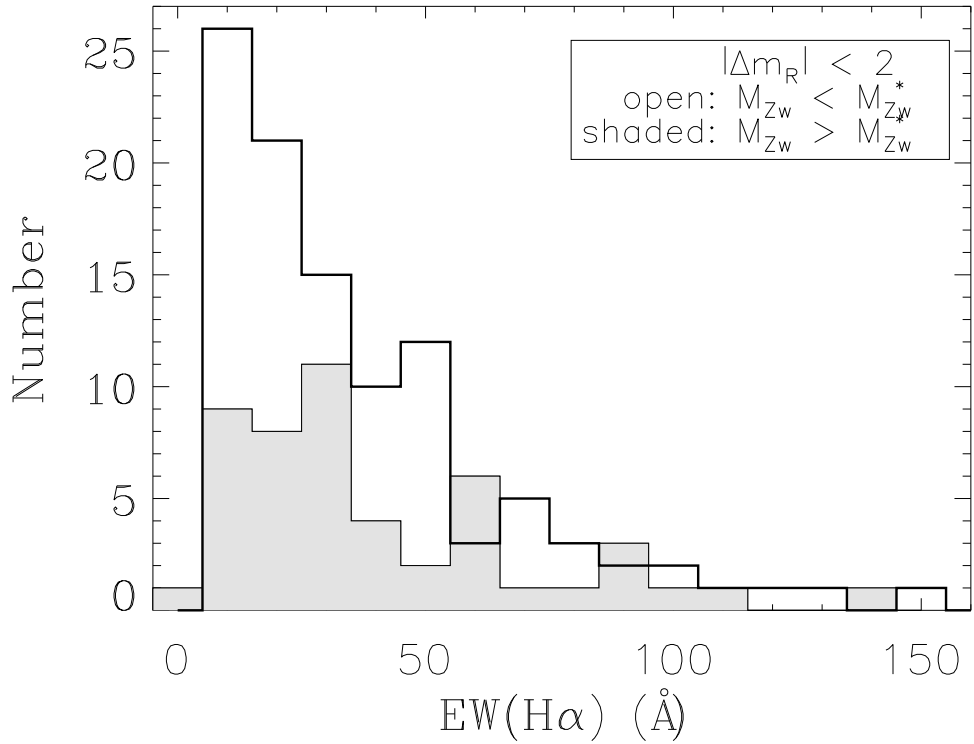}
\caption{
The distributions of EW(H$\alpha$) for the $M_{Zw} > M_{Zw}^{*}$ 
and $M_{Zw} < M_{Zw}^{*}$ galaxies in the DW+EB sample.
The $1 \sigma$ fractional error in EW(H$\alpha$) is $18\%$, 
and $\pm 0.41$ mag for $M_{Zw}$.  Not shown are the 
EW(H$\alpha$)$ < 3.5$~\AA\ (corrected for stellar absorption)
galaxies, which make up $21\%$ (13/62) of the 
$M_{Zw} > M_{Zw}^{*}$ sample and $27\%$ (38/141) of the 
$M_{Zw} < M_{Zw}^{*}$ galaxies.}  % updated 7/8
\label{ew-abslt2}
\end{center}
\end{figure}

The set of 203 galaxies with $\left | \Delta m_R \right | < 2$ contains
141 galaxies ($69\%$) with $M_{Zw} < M_{Zw}^{*}$ and 62 galaxies ($31\%$) 
with $M_{Zw} > M_{Zw}^{*}$.  A K-S test of the distributions of 
EW(H$\alpha$) in the two groups shows that the distributions are 
similar: the probability of their deriving from the same parent sample 
is $31\%$ (see Fig.~\ref{ew-abslt2}).

\begin{figure}[htb!]
\begin{center}
\includegraphics[width=4.5in,angle=90]{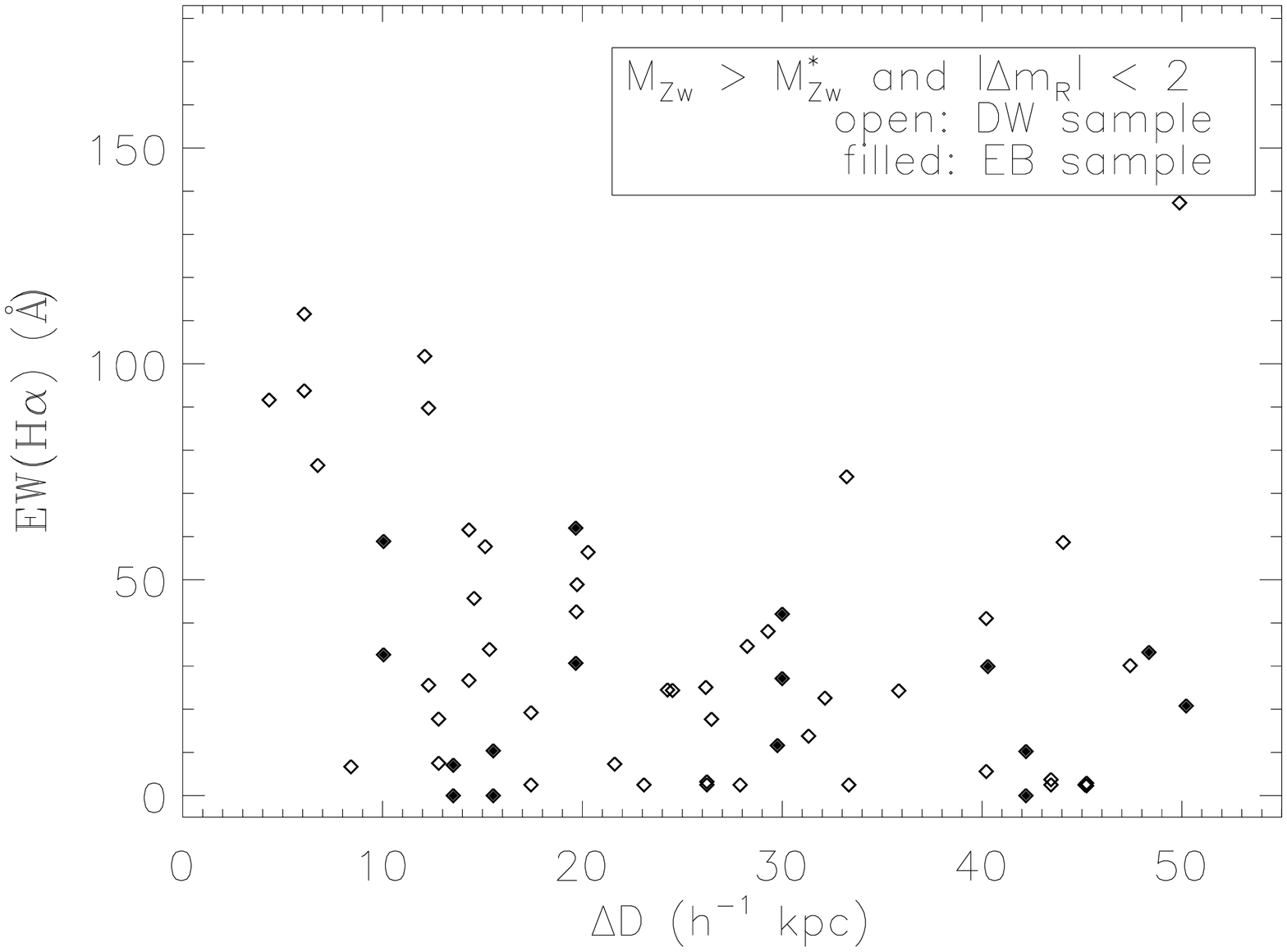}
\caption{$\Delta D$ versus EW(H$\alpha$)  for the galaxies with magnitude
$M_{Zw} > M_{Zw}^{*}$.  The $1 \sigma$ fractional error is $18\%$ in
EW(H$\alpha$), and $\pm0.41$ mag for $M_{Zw}$.}
\label{ew-sep-18a}
\end{center}
\end{figure}

\begin{figure}[htb!]
\begin{center}
\includegraphics[width=4.5in,angle=90]{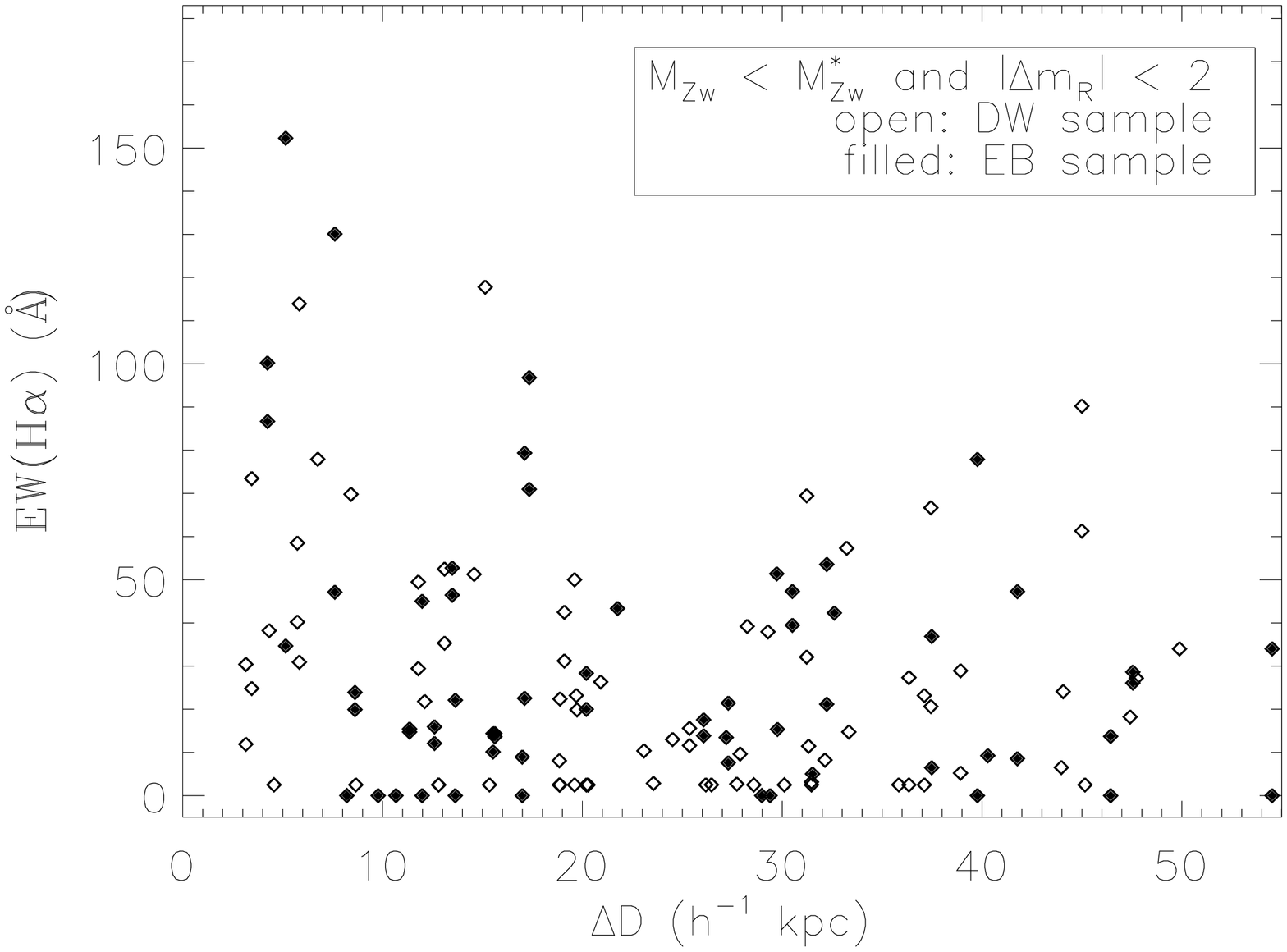}
\caption{$\Delta D$ versus EW(H$\alpha$) for the galaxies with magnitude
$M_{Zw} < M_{Zw}^{*}$.  The $1 \sigma$ fractional error is $18\%$ in
EW(H$\alpha$), and $\pm0.41$ mag for $M_{Zw}$.}
\label{ew-sep-18b}
\end{center}
\end{figure}

Both subsets of galaxies with $M_{Zw} > M_{Zw}^{*}$ and with
$M_{Zw} < M_{Zw}^{*}$, where $\left | \Delta m_R \right | < 2$,
demonstrate a probable correlation between EW(H$\alpha$) and $\Delta D$. 
The more luminous galaxies, $M_{Zw} < M_{Zw}^{*}$, show a 
correlation of $C_{SR} = -0.14$ with $P_{SR} = 9.8\times10^{-2}$.  The low 
luminosity galaxies, $M_{Zw} > M_{Zw}^{*}$, show 
$C_{SR} = -0.32$ with $P_{SR} = 1.2\times10^{-2}$.  
Both subsets show a similar trend in the correlation between 
EW(H$\alpha$) and $\Delta D$: higher values of EW(H$\alpha$) 
correlate with small spatial separations 
(Figs.~\ref{ew-sep-18a} and \ref{ew-sep-18b}).   
Our results suggest that the intrinsic luminosity of the galaxy 
plays little role in the effectiveness of the tidally triggered 
star formation induced by its companion galaxy.

\subsection{Relative Magnitude and EW(H$\alpha$)} \label{res-rel}

\begin{figure}[htb!]
\begin{center}
\includegraphics[width=4.5in,angle=90]{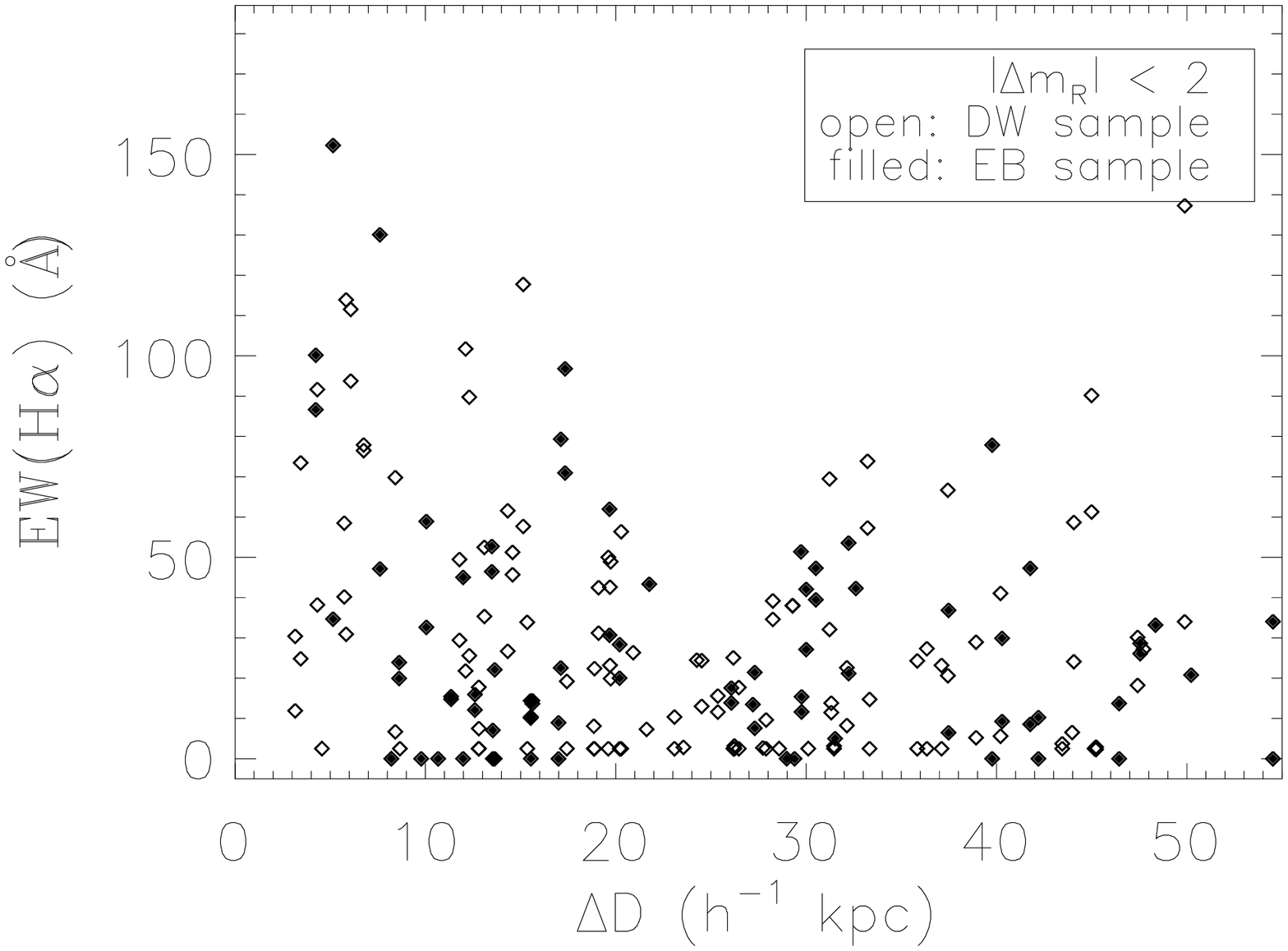}
\caption{$\Delta D$ versus EW(H$\alpha$)  for the galaxies with
$\left | \Delta m_R \right | < 2$.  The $1\sigma$ fractional error is 
$18\%$ for EW(H$\alpha$), and $0.17$ mag for $\left | \Delta m_R \right |$.}
\label{ew-sep-delmra}
\end{center}
\end{figure}

\begin{figure}[htb!]
\begin{center}
\includegraphics[width=4.5in,angle=90]{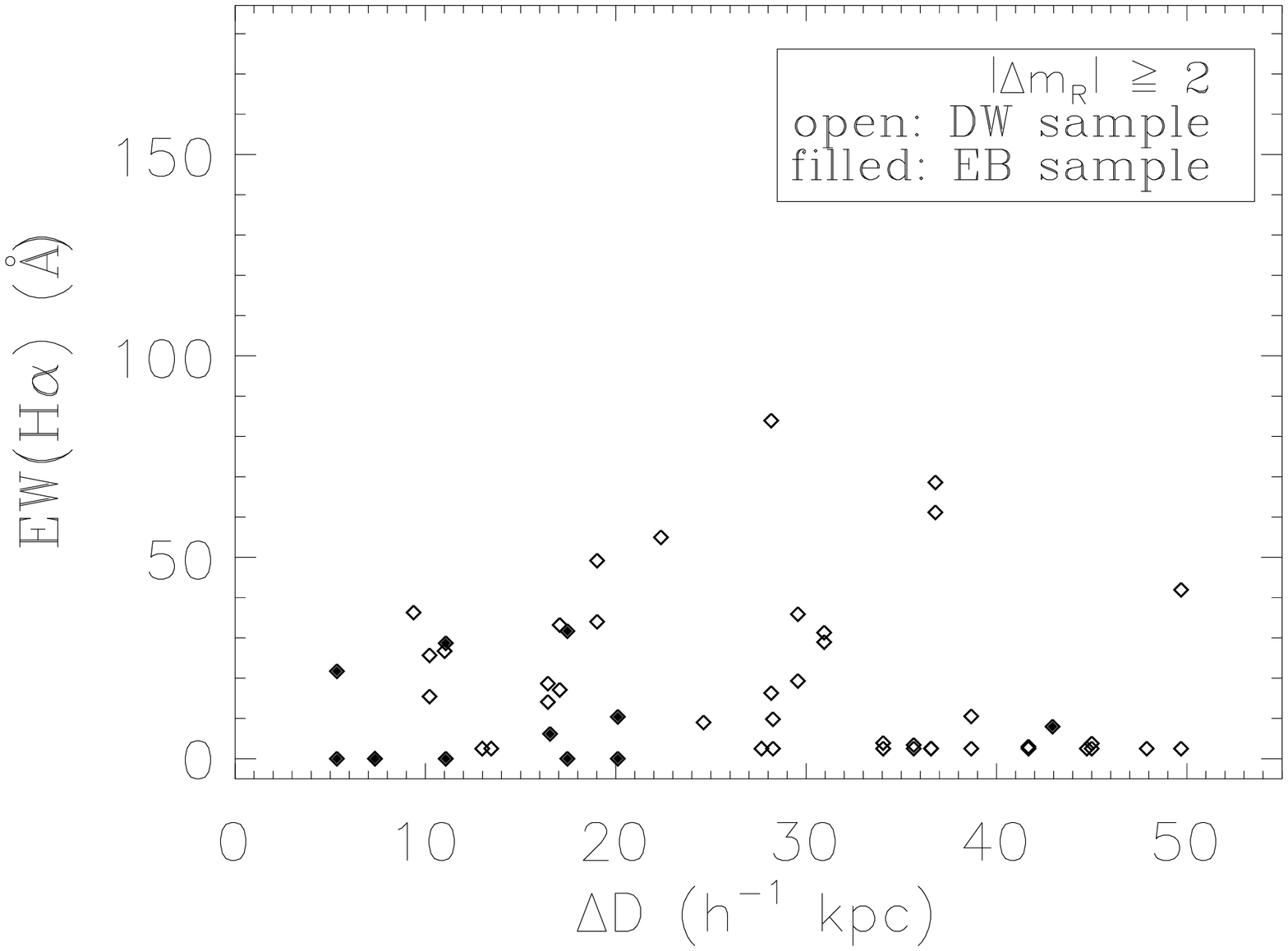}
\caption{$\Delta D$ versus EW(H$\alpha$)  for the galaxies with
$\left | \Delta m_R \right | \geq$.  The $1\sigma$ fractional error is 
$18\%$ for EW(H$\alpha$), and $0.17$ mag for $\left | \Delta m_R \right |$.}
\label{ew-sep-delmrb}
\end{center}
\end{figure}

Next we examine the effect of the relative magnitude of the pair on the
correlation between EW(H$\alpha$) and $\Delta D$.  
Figs.~\ref{ew-sep-delmra} and \ref{ew-sep-delmrb}
show EW(H$\alpha$) versus $\Delta D$ for the 260 galaxies with FLWO 1.2~m
photometry, grouped by magnitude difference between the galaxy 
and its nearest neighbor.  We separate the galaxies into two subsets,
according to $\left | \Delta m_R \right |$.  The 57 individual galaxies 
with $\left | \Delta m_R \right | \geq 2$ include 26 galaxies ($46\%$) with
$M_{Zw} < M_{Zw}^{*}$, and 31 galaxies ($54\%$) with $M_{Zw} > M_{Zw}^{*}$.  
(Note that the number of individual galaxies with 
$\left | \Delta m_R \right | \geq 2$ can be odd because our sample 
includes multiplets, in which the number of galaxies in the compact 
group $> 2$.  Each galaxy is compared to its nearest neighbor.)
All of the 57 galaxies with $\left | \Delta m_R \right | \geq 2$,
regardless of $M_{Zw}$, are included in this study of the effect of 
the relative magnitude of the pair because we showed in \S\ref{absolutemag} 
that intrinsically luminous and low luminosity galaxies exhibit similar
trends in the correlation of EW(H$\alpha$) and $\Delta D$.
We choose the boundary of $\left | \Delta m_R \right | = 2$ because it 
corresponds to a mass ratio of $\sim 10$ and allows us to probe a region 
not well covered in other studies.  This division provides a decent sample 
size with $\left | \Delta m_R \right | \geq 2$.  Changing the boundary by 
$\pm 0.3$~mag does not qualitatively affect the results.

Applying the Spearman rank test,
we find a clear correlation between EW(H$\alpha$) and $\Delta D$
for the 203 galaxies with $\left | \Delta m_R \right | < 2$,
where $C_{SR} = -0.18$ and $P_{SR} = 8.9\times10^{-3}$. 
The Spearman rank test measures no correlation between 
 EW(H$\alpha$) and $\Delta D$ for the 57 galaxies with 
$\left | \Delta m_R \right | \geq 2$.
The absence of galaxies with EW(H$\alpha) \gtrsim 70$~\AA\ is
evident for the pairs with $\left | \Delta m_R \right | \geq 2$
 (Fig.~\ref{ew-sep-delmrb}).  We note, however, that a larger
sample of $\left | \Delta m_R \right | \geq 2$ pairs at separations
$\Delta D < 5~h^{-1}$~kpc would be helpful for verifying this result.

\begin{figure}[htb!]
\begin{center}
\includegraphics[width=4.5in,angle=90]{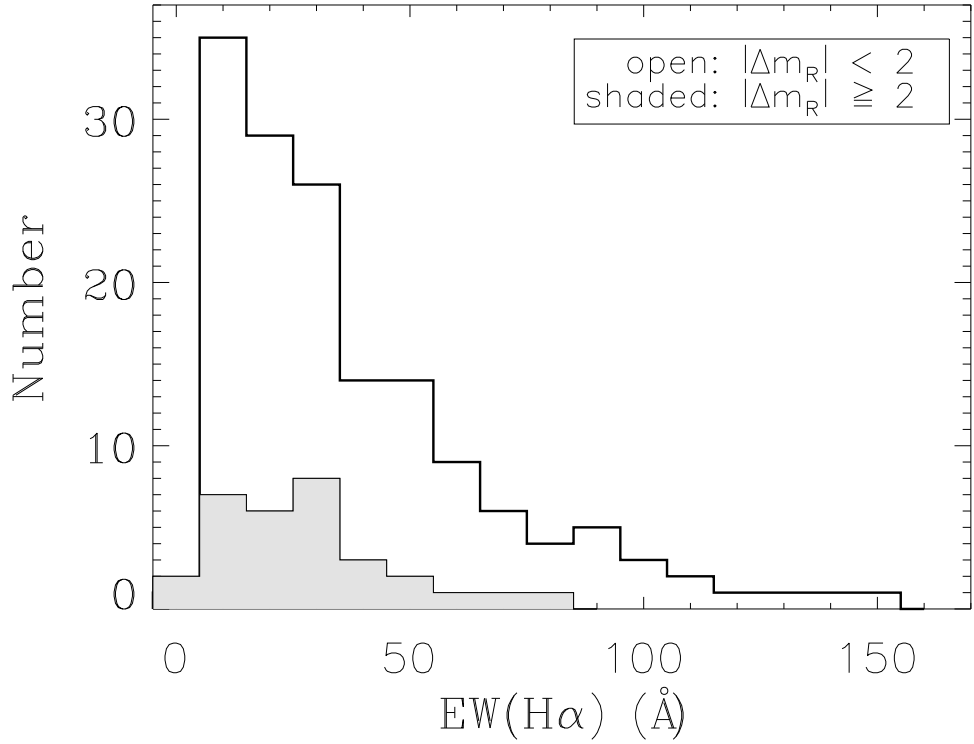}
\caption{Distribution of EW(H$\alpha$) for the 
$\left | \Delta m_R \right | < 2$ and $\left | \Delta m_R \right | \geq~2$ 
subsets.   Note the absence of high EW(H$\alpha$) for the 
$\left | \Delta m_R \right | \geq 2$ galaxies.  Values of 
EW(H$\alpha$) $< 3.5$~\AA\ (corrected for Balmer absorption) are 
excluded from the plot.  The EW(H$\alpha$) $< 3.5$~\AA\ make up 
$25\%$ (51 out of 203) of the $\left | \Delta m_R \right | < 2$ sample 
and $46\%$ (26 out of 57) of the $\left | \Delta m_R \right | > 2$ sample.}
\label{ew-gtlt}
\end{center}
\end{figure}

Comparison of the distributions of EW(H$\alpha$) provides
another test of whether the galaxies in pairs with large or small 
luminosity contrast are similarly affected by tidal interactions.
Fig.~\ref{ew-gtlt} shows the distributions  of EW(H$\alpha$) for 
galaxies with $\left | \Delta m_R \right | < 2$ and galaxies with 
$\left | \Delta m_R \right | \geq 2$.  The K-S 
probability of the two distributions deriving from the same parent
sample is  $1.5\times10^{-4}$.  The galaxies are unlikely to 
be drawn from the same parent sample.  This result suggests that
luminosity contrast of the pair influences the strength or age of
the tidally triggered star formation.

In addition to testing the influence of the relative luminosity, 
$\left | \Delta m_R \right |$, on the correlation between EW(H$\alpha$) 
and $\Delta D$, we also test the correlation between EW(H$\alpha$) and 
$\left | \Delta m_R \right |$ directly.  Fig.~\ref{ew-magdifR} shows 
the relationship between  EW(H$\alpha$) and $\left | \Delta m_R \right |$.  
A Spearman rank test of EW(H$\alpha$) versus $\left | \Delta m_R \right |$ for 
all 260 galaxies with FLWO 1.2~m photometry measures $C_{SR} = -0.13$ 
with $P_{SR} = 3.6 \times10^{-2}$.  The negative correlation coefficient 
indicates that pairs with similar magnitudes, small 
$\left | \Delta m_R \right |$, have the largest EW(H$\alpha$).  The 
galaxies in low luminosity contrast systems, regardless of their intrinsic 
luminosity, are more strongly affected by the tidal interaction.  

At $\Delta D < 5~h^{-1}$~kpc, there is an absence of pairs with 
$\left | \Delta m_R \right | \geq 2$ and very few pairs with
$\left | \Delta m_R \right |  2$ (see Figs.~\ref{ew-sep-delmra} and \ref{ew-sep-delmrb}).
We have an observational bias against identifying very faint companions
close to bright galaxies.  The faintest galaxies in the DW sample
result from our visual identification of companions around UZC galaxies
(\S\ref{pairs-sample}).  Furthermore, low luminosity companions
may be tidally disrupted by the primary galaxy before reaching 
$\Delta D \sim 0$ \citep{hern_mihos95}.  It is possible
that we fail to observe any galaxies  with 
$\left | \Delta m_R \right | \geq 2$ and $\Delta D < 5~h^{-1}$~kpc because
they have been disrupted.  It is also possible that we cannot detect
undisrupted faint objects against the brighter primary.

\section{Discussion} \label{discussion}
\setcounter{footnote}{1}

\begin{figure}[htb!]
\begin{center}  
\includegraphics[width=4.5in,angle=90]{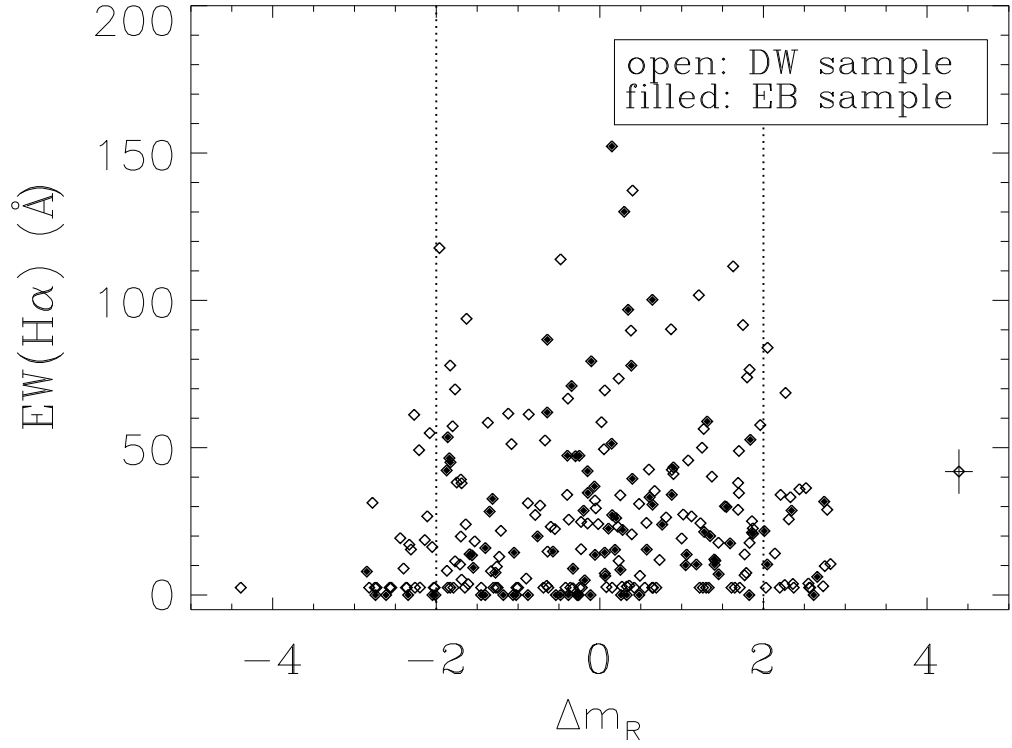}
\caption{$\Delta m_R$ versus EW(H$\alpha$) for the DW+EB sample.
The brighter of the pair has $\Delta m_R < 0$, and the fainter 
of the pair has $\Delta m_R > 0$.  Representative error
bars of $\pm 0.17$ mag in $\Delta m_R$ and $\pm 18\%$ in
EW(H$\alpha$) are shown on the right most point.}
\label{ew-magdifR}
\end{center}
\end{figure}

The subset of 57 galaxies ($22\%$) of our pairs sample with 
$\left | \Delta m_R \right| \geq 2$ 
provides an opportunity to extend the study of tidally triggered star 
formation to minor interactions.  The normalized star formation rate for 
our galaxies, measured in terms of EW(H$\alpha$), depends strongly on the 
relative luminosity of the galaxies for the $\left | \Delta m_R \right| \geq 2$
subset.  The highest values of EW(H$\alpha$) occur for the galaxies in
pairs of similar luminosity, where $\left | \Delta m_R \right | \sim 0$ 
(Fig.~\ref{ew-magdifR}).

The normalized star formation rate as a function of projected
spatial separation for the minor encounters differs from that of the 
major encounters.  The galaxies with $\left | \Delta m_R \right| <2$ 
show a clear correlation ($P_{SR} = 8.9\times10^{-3}$)
between EW(H$\alpha$) and the projected spatial 
separation, $\Delta D$; the $\left | \Delta m_R \right| \geq 2$ galaxies 
do not. A larger sample is needed to probe the response to tidal 
interactions for the brighter and the fainter 
of the $\left | \Delta m_R \right | \geq 2$ galaxies separately.

The correlations we find between EW(H$\alpha$) - $\Delta D$ 
and EW(H$\alpha$) - $\Delta V$ for our sample as a whole are in
good agreement with the results of \citet{bgk}\footnotemark[1]. 
A study by \citet{nikolic04} likewise shows an increase in the
specific star formation rate at small projected separations
$< 30$~kpc, in their sample of 12,492 SDSS galaxies with $M_r < -20.45$.  
They detect a correlation between specific star formation rate and 
projected separation out to 300~kpc for late-type galaxies.
Nikolic et al. also find that the specific star formation rate 
decreases for pairs with increasing recessional velocity differences.
Similarly, \citet{lambas03}  find that their 1258 field galaxy pairs
from the 2dF Survey with $z \leq 0.1$ exhibit enhanced star formation for 
$\Delta D < 25 h^{-1}$~kpc and $\Delta V < 100$~km s$^{-1}$.
The work of \citet{hern05} further supports the correlation between
projected separation of pair galaxies and their star formation rates.
Hern$\acute{a}$ndez-Toledo et al. study the light concentration $C$, asymmetry $A$,
and clumpiness $S$ of 66 disk galaxies in spiral-spiral pairs
\citep{hern01,karach72}, compared to a set of 113 non-interacting galaxies 
and 66 ultraluminous infrared galaxies (ULIRGs; \citealp{conselice03}), which are associated with recent
interactions.  They conclude that the $CAS$ parameters of the closest pairs 
are similar to those of the ULIRGs', while the $CAS$ parameters of the widest 
pairs are more similar to the isolated galaxy sample.  
By contrast, \citet{donzelli-p97} do not find a significant correlation
between EW(H$\alpha$ + [N\textsc{II}]) and the projected distance
between pair galaxies in their study of 27 physical pairs.  Their small
sample size and possible selection effects make it difficult to 
evaluate their results.

The numerical simulations of \citet{perez05} support the connection
between enhanced star formation and the proximity of galaxies in
pairs.  Their simulated catalog includes galaxies in pairs
(three dimensional separation $r < 100~h^{-1}$~kpc) and galaxies without a 
close companion formed in a $\Lambda$CDM cosmology.  
 They find that galaxies with a companion closer than
$30 \pm 10~h^{-1}$~kpc demonstrate an excess of star formation
activity compared to galaxies without a close companion.
However, not all pair galaxies have
enhanced star formation: $40\%$ of the simulated galaxy pairs
with a companion closer than 30 $h^{-1}$~kpc do not.  
The availability of gas, the depth of the potential well, and 
the physical separation may help determine the tidally
driven gaseous inflow that triggers the burst
\citep{barnes96,tissera00,perez05}.  Perez et al.'s analysis
of the simulated catalog in two dimensional projection yields consistent
results for enhanced star formation for galaxies with a
companion at projected separation $r_p < 25~h^{-1}$~kpc.

\footnotetext{Our sample derives from the same parent sample as
that of \citet{bgk}.  The samples differ in that our sample includes 
pairs where the primary member is a UZC galaxy and the companion
is identified by follow-up observations, while the sample of 
\citet{bgk} includes only pairs where both members are UZC galaxies.  
See \S\ref{pairs-sample} for our
sample selection and \S\ref{character} for the sample characteristics.}

We draw similar conclusions on the effects of absolute luminosity on 
measured star formation rates to those reported by \citet{lambas03} in 
their study of 2dF field galaxy pairs.  In our sample, subsets of 
intrinsically luminous ($M_{Zw} < M_{Zw}^{*}$) galaxies and low luminosity 
galaxies ($M_{Zw} > M_{Zw}^{*}$) show similar correlations between 
EW(H$\alpha$) and  $\Delta D$. 
Although Lambas et al.'s method for measuring enhanced star formation 
differs from ours, their study similarly shows that the intrinsic luminosity 
of the pair galaxy has no effect on the mean star formation excess when 
compared to isolated galaxies, although they find that low luminosity 
galaxies have higher absolute mean stellar birthrate $b$ parameters.

We compare the effects of relative luminosity on star formation
activity in our sample with the results of other recent studies.
\citet{nikolic04} find  no dependence on the mass ($z$-band magnitude)
or morphological type (concentration index) of the companion galaxy
in their sample of SDSS close pairs.
Nikolic et al. examine the distributions of specific star formation 
rates for subsets of relative z-band magnitude, $-2 < \Delta m_z \leq -1$, 
$-1 < \Delta m_z \leq 0$, and $0 < \Delta m_z <2$, and 
find no evidence for a difference in the distributions at the $50\%$
confidence level.    If we divide our sample 
into similar bins by $\Delta m_R$, we find that the distributions of 
EW(H$\alpha$) for galaxies with $-2 < \Delta m_R \leq -1$ and 
$-1 < \Delta m_R \leq 0$ have a $33\%$ probability of deriving from 
the same parent sample, and the galaxies with  $0 < \Delta m_R \leq 2$ 
have a $24\%$  probability of deriving from the same parent sample 
as the $-2 < \Delta m_R \leq 0$ galaxies.  These similarities in 
distributions of EW(H$\alpha$) are consistent with the results of 
Nikolic et al.

Extending the comparison of distributions of EW(H$\alpha$)
for subsets of luminosity contrast beyond those considered by \citet{nikolic04}
reveals a different story.  Our sample shows that the
distribution of EW(H$\alpha$) for the 
$\left | \Delta m_R \right | \geq 2$ galaxies has only a 
$0.02\%$ probability of deriving from the same parent sample as
the $\left | \Delta m_R \right | < 2$ galaxies.  Our results suggest 
that the luminosity ratio (mass ratio) of the pair does influence 
the effectiveness of the tidally triggered star formation.  The
effect becomes apparent only when the luminosity contrast is large.

We find that galaxies in major interactions are more likely to show 
enhanced star formation activity than galaxies in minor interactions.
\citet{lambas03} describe the same general trend in their study of 2dF 
galaxies: they find that galaxy pairs 
of similar luminosity, defined as $L_1/L_2 < 0.5$ ($\Delta m < 0.75$), 
reveal enhanced star formation in both members.  Their galaxy pairs of 
dissimilar luminosity, $L_1/L_2 > 0.5$, show less star formation 
enhancement than pairs of similar luminosity.  In our data, 
Fig.~\ref{ew-magdifR} clearly shows a peak in EW(H$\alpha$) around
$\Delta m_R = 0$, and decreases for galaxies with larger 
magnitude differences.  A Spearman rank correlation test of
EW(H$\alpha$) and $\left | \Delta m_R \right |$ shows that
high values of EW(H$\alpha$) correlate with small magnitude 
differences (i.e. nearly equal luminosities).

In a related field, \citet{dasyra06} study the context for 
ULIRG activity in galaxy merger remnants.  Their analysis includes
23 ULIRGs in binary merger remnants that still have two distinct nuclei.  
  Most of the ULIRGs in their sample
are triggered by interactions between galaxies of nearly
equal mass.  The average mass ratio of the pair is 1.5:1.  Although
some of their pairs have a mass ratio of 3 to 1, Dasyra et al. find
that galaxy pairs with larger ratios do not produce ULIRGs.  Because
ULIRGs occur when gas-rich, disk galaxies merge 
\citep[e.g. ][]{downes98,bryant99}, we compare the properties of 
the ULIRG host galaxies with our interacting pairs.  Our results
are consistent in that pairs with small magnitude differences
(i.e. similar mass) appear to trigger central star formation more 
effectively than pairs with large magnitude differences.

\section{Conclusions} \label{conclusion}

We assemble a sample of 345 galaxies in 167 pairs and compact groups
to measure the star formation activity as a function of intrinsic and
relative properties of the galaxies.  Our sample derives from the
CfA2 Redshift Survey pairs sample (see \citealt{geller-huchra89} for 
a description of the CfA2 Redshift Survey; \citealt{bgk,barton01} 
for the CfA2 pairs sample).  We construct our sample with the aim
of including pairs of dissimilar luminosity because minor interactions
are important for galaxy formation in the hierarchical formation model
\citep{somerville_prim99,kauffmann99a,kauffmann99b,diaferio99}.
Our sample contains $22\%$ of the pairs with photometry with 
$\left | \Delta m_R \right | \geq 2$.

To isolate the intrinsic  galaxy properties from the properties of the 
interaction that influence the effectiveness of the tidally triggered star 
formation, we examine the  EW(H$\alpha$) - $\Delta D$ correlation and
the distributions of EW(H$\alpha$) for various subsets of our sample.
We find that:

\begin{enumerate}

\item Galaxies with $M_{Zw} < M_{Zw}^{*}$, and galaxies with 
$M_{Zw} > M_{Zw}^{*}$ show a correlation between EW(H$\alpha$) 
and $\Delta D$, provided $\left | \Delta m_R \right | < 2$.

\item The distribution of EW(H$\alpha$) for the $M_{Zw} > M_{Zw}^{*}$ 
galaxies is similar to the distribution of EW(H$\alpha$) for the
$M_{Zw} < M_{Zw}^{*}$ galaxies, again provided 
$\left | \Delta m_R \right | < 2$.

\item Galaxies in pairs of small luminosity contrast, $\left | \Delta m_R
\right | < 2$, show a strong correlation between EW(H$\alpha$)
and $\Delta D$.

\item Galaxies in pairs of large luminosity contrast, 
$\left | \Delta m_R \right | \geq 2$, show no significant correlation 
between  EW(H$\alpha$) and $\Delta D$.

\item The distribution of EW(H$\alpha$) for the  $\left | \Delta m_R
\right | < 2$ galaxies differs significantly from the distribution
of EW(H$\alpha$) for the  $\left | \Delta m_R \right | \geq 2$ galaxies.
Very few galaxies with  $\left | \Delta m_R \right | \geq 2$ have 
EW(H$\alpha) > 70$~\AA, in contrast to the  $\left | \Delta m_R
\right | < 2$ galaxies.

\item The largest values of EW(H$\alpha$) are associated with galaxies 
in pairs of $\left | \Delta m_R \right | \sim 0$.

\end{enumerate}

The relative luminosity (and thus presumably mass) of the companion 
galaxy is more important in a gravitational tidal interaction than the 
intrinsic luminosity of the galaxy.  Galaxies in pairs of similar 
luminosity are more strongly affected by tidally triggered star formation 
than galaxies in pairs with $\left | \Delta m_R \right | \geq 2$.  

Not all galaxies in the pairs sample exhibit significant star formation: 
$32\%$ (74 out of 230) of the DW sample has EW(H$\alpha$) $< 3.5$~\AA\ (corrected 
for Balmer absorption).  Some galaxies fail to respond to gravitational
tidal forces because they lack available gas \citep[e.g.][]{barnes96}, 
and other pairs are merely superpositions.  The pair galaxies may just be 
starting to approach each other for the first time and have not yet 
experienced a close pass in their orbits.  Galaxy structure and
orbital geometry influence the effectiveness of the tidally triggered
star formation \citep{mihos_hern96}. In addition, the lowest 
mass galaxies could be strongly affected by other energetic processes 
such as supernova triggered star formation \citep{lada78,elmegreen95,boss03}
and could show enhanced star formation activity independent of pair separation.  

Our observed correlation between EW(H$\alpha$) and $\Delta D$ is
consistent with the theoretical interpretation that tidally
triggered star formation results from gas driven to the center of the
galaxy by tidal interactions just after perigalacticon, disrupting the 
system and causing a burst of star formation \citep{mihos_hern96}.
The absence of  $\left | \Delta m_R \right | \geq 2$ galaxies with 
values of EW(H$\alpha) \gtrsim 70$~\AA\ suggests that the relative mass 
of the galaxies influences the effectiveness of tidally triggered star 
formation.  A stronger test of triggered star formation in minor
interactions would include more $\left | \Delta m_R \right | \geq 2$ 
pairs at physical separation $\Delta D < 5~h^{-1}$~kpc, which are
an observational challenge.

\acknowledgments
We thank Daniel Fabricant, Warren Brown, and Lisa Kewley for their 
insights and assistance throughout this project.  We thank Perry Berlind
and Mike Calkins for taking the FAST spectra, and we thank Susan Tokarz
for her work in reducing the spectra.  We also enjoyed numerous
conversations with Jenny Greene, Scott Kenyon, and Michael Kurtz.
We thank the referee for helpful comments that strengthened the 
paper and prompted the comparison with the 15R-North Survey.
This project was supported in part by the Smithsonian Institution, 
Harvard University, and the University of California, Irvine.

\clearpage

% figures

% tables

\clearpage

\begin{deluxetable}{lllllll} 
\tabletypesize{\scriptsize}
\tablecolumns{7} 
\tablewidth{0pc} 
\tablecaption{Galaxy data for members of pairs.\tablenotemark{*}  
\label{pairs-table} }
\tablehead{ \colhead{ID} & \colhead{RA (J2000)} & \colhead{Dec (J2000)} & 
\colhead{$cz$\tablenotemark{a}} & \colhead{EW(H$\alpha$)\tablenotemark{b}} & \colhead{$\Delta m_R$\tablenotemark{c}} & \colhead{Type\tablenotemark{d}}\\
 & $\fh \ \ \fm \ \ \fs$ & $\degr \ \ \arcmin \ \ \arcsec$ & (km s$^{-1}$) & (\AA) &}
\startdata 
000348+17090\_a & 0 6 21.74 & 17 26 15.49 & $5770 \pm2$ & 49 & -0.3 & gal\\ 
000348+17090\_b & 0 6 21.44 & 17 25 47.43 & $5589 \pm1$ & 132 & 0.3 & gal\\ 
081648+22120\_a & 8 19 48.49 & 22 1 57.68 & $3476 \pm9$ & 2 & -2.7 & gal\\ 
081648+22120\_b & 8 19 41.34 & 22 2 30.30 & $3354 \pm4$ & 33 & 2.7 & gal\\ 
115454+32360\_a & 11 57 31.62 & 32 20 27.78 & $3259 \pm6$ & 154 & 0.1 & gal\\ 
115454+32360\_b & 11 57 43.95 & 32 17 39.62 & $3324 \pm9$ & 81 & -0.1 & gal\\ 
120306+09160\_a & 12 5 36.29 & 8 59 15.90 & $6309 \pm9$ & 23 & 1.3 & gal\\ 
120306+09160\_b & 12 5 42.43 & 8 59 22.53 & $6230 \pm11$ & 9 & -1.3 & gal\\
134412+44050\_a & 13 46 23.68 & 43 52 18.60 & $2490 \pm15$ & 2 & -2.3 & gal\\ 
134412+44050\_b & 13 46 18.53 & 43 51 3.70 & $2360 \pm3$ & 30 & 2.3 & gal\\ 
\enddata  
\tablenotetext{*}{Full table appears in electronic edition.}
\tablenotetext{a}{Error measurement is based on the
derived $r$ value \citep{td79} and the FWHM of the correlation peak used to 
obtain $cz$.  See \S3.2 in \citet{kurtz_mink} for a detailed description 
of the error measurement.}
\tablenotetext{b}{EW(H($\alpha$) is corrected for Balmer absorption. 
Error in EW(H$\alpha$) is $\pm18\%$, described in \S\ref{spec-sample}.}
\tablenotetext{c}{$\Delta m_R < 0$ for the brighter of the pair and
$\Delta m_R > 0$ for the fainter.}
\tablenotetext{d}{Object type: galaxy (gal), intermediate (int), or AGN.}
\end{deluxetable}

\begin{deluxetable}{lllll} 
\tablecolumns{5} 
\tablewidth{0pc} 
\tablecaption{Galaxy data for non-member galaxies.\tablenotemark{*}  
\label{isolated} }
\tablehead{ \colhead{ID} & \colhead{RA (J2000)} & \colhead{Dec (J2000)} & 
\colhead{cz\tablenotemark{a}} \\
 &  $\fh \ \ \fm \ \ \fs$ & $\degr \ \ \arcmin \ \ \arcsec$ & (km s$^{-1}$)}
\startdata 
001412+06470\_b & 0 16 38.79 & 7 6 55.33 & $11851 \pm20$ \\ 
002018+06330\_b & 0 22 53.52 & 6 49 38.22 & $15100 \pm20$ \\ 
003648+36050\_b & 0 39 33.52 & 36 23 33.42 & $16120 \pm16$ \\ 
004154+16320\_c & 0 44 33.32 & 16 50 12.59 & $23843 \pm17$ \\ 
004212+04530\_b & 0 44 57.93 & 5 8 44.96 & $38646 \pm31$ \\ 
004336+19130\_b & 0 46 8.60 & 19 31 16.70 & $29611 \pm28$ \\ 
004336+19130\_e & 0 46 12.54 & 19 33 20.50 & $28511 \pm30$ \\ 
004736-02120\_b & 0 50 5.55 & -1 55 59.47 & $24282 \pm18$ \\ 
005706+17450\_b & 0 59 40.29 & 17 58 25.31 & $26257 \pm17$ \\ 
010218+04300\_b & 1 4 44.73 & 4 47 27.14 & $13754 \pm16$ \\ 
\enddata  
\tablenotetext{*}{Full table appears in electronic edition.}
\tablenotetext{a}{Error measurement is based on the
derived $r$ value \citep{td79} and the FWHM of the correlation peak used to 
obtain $cz$.  See \S3.2 in \citet{kurtz_mink} for a detailed description 
of the error measurement.}
\end{deluxetable}

\begin{deluxetable}{llllll} 
\tablecolumns{6} 
\tablewidth{0pc} 
\tablecaption{Absolute photometry for galaxies in DW sample.
\label{absolute-table} }
\tablehead{ \colhead{ID} & \colhead{RA (J2000)} & \colhead{Dec (J2000)} & \colhead{$M_R$\tablenotemark{a}} & 
\colhead{$M_B$\tablenotemark{a}} & \colhead{Obs. Date} \\
 &  $\fh \ \ \fm \ \ \fs$ & $\degr \ \ \arcmin \ \ \arcsec$  & & & }
\startdata 
142812+00280\_a & 14 30 45.92 & 0 14 52.39 &  -23.59 & -21.51 & 2003 Jun 01  \\ 
142812+00280\_b & 14 30 43.03 & 0 15 11.12  & -23.20 & -21.15  & 2003 Jun 01  \\ 
143412+02360\_a & 14 36 41.93 & 2 23 9.61 & -21.34 & -19.34    & 2003 Jun 01  \\ 
143412+02360\_b & 14 36 41.43 & 2 22 26.25 & -20.26 &  -18.22  & 2003 Jun 01  \\  
150200+42180\_a & 15 3 50.43 & 42 6 56.10 & -22.47  &  -20.00 &   2003 Jun 01 \\ 
150200+42180\_b & 15 3 39.71 & 42 7 34.45 & -20.60  & -18.28  &  2003 Jun 01
\enddata  
\tablenotetext{a}{Error in $M_B$ and $M_R$ is 0.12~mag.}
\end{deluxetable} 

\begin{deluxetable}{lllll} 
\tabletypesize{\scriptsize}
\tablecolumns{5} 
\tablewidth{0pc} 
\tablecaption{Comparison of EW(H$\alpha$) in different samples. 
\label{comp-table} }
\tablehead{\colhead{Sample} & \colhead{$> 10$~\AA} & \colhead{$> 40$~\AA} & \colhead{$> 70$~\AA} & \colhead{Num. Gal.}}
\startdata 
DW  & $59\%$ & $23\%$ & $9\%$ & 230\\
15R-North\tablenotemark{a} & $25\%$ & $4\%$ & $0.2\%$  & 1675\\
EB  & $66\%$ & $23\%$ & $9\%$ & 92\\
UZC\tablenotemark{b} &  $36\%$ &  $6\%$ &  $2\%$ & 12562 \\
\enddata  
\tablenotetext{a}{Galaxies selected to match redshift range and apparent
magnitude range of DW sample: $2300 < cz < 16,500$~km~s$^{-1}$, and $m_R < 17$.
Galaxies with AGN-like spectra are excluded.}
\tablenotetext{b}{Galaxies selected with $2300 < cz < 16500$~km~s$^{-1}$, and $m_{Zw} < 18$.}
\end{deluxetable} 

\end{document}